\documentclass[journal]{IEEEtran}

\usepackage{epsfig}
\usepackage{amsmath}
\usepackage{amssymb}
\usepackage{amsthm}
\usepackage{multirow}
\usepackage{graphicx} 
\usepackage{blindtext}
\usepackage{xcolor}

\theoremstyle{definition}

\newcommand{\f}{\operatorname}

\ifCLASSINFOpdf
\else
\fi

\begin{document}

\title{Multiple repairable systems under dependent competing risks with nonparametric Frailty}

\author{Marco Pollo~Almeida, 
        Rafael~Paix\~ao, Pedro~Ramos, Vera~Tomazella,
         Francisco Louzada and Ricardo~Ehlers
\thanks{Marco Pollo Almeida, Pedro L. Ramos and Francisco Louzada work at the Institute of Mathematical Science and Computing at the University of S\~ao Paulo - USP, S\~ao Carlos,
SP, Brazil, e-mail: mpa@usp.br, pedrolramos@usp.br, louzada@icmc.usp.br. Vera Tomazella works at the Department
of Statistics, UFSCar, S\~ao Carlos,
SP, Brazil, e-mail: vera@ufscar.br.}
\thanks{Manuscript received November 1.}}

\markboth{IEEE Transactions on Reliability,~Vol.~00, No.~0, November~2019}%
{Shell \MakeLowercase{\textit{et al.}}: Bare Demo of IEEEtran.cls for IEEE Journals}

\maketitle

\begin{abstract}
The aim of this article is to analyze data from multiple repairable systems under the presence of dependent competing risks. In order to model this dependence structure, we adopted the well-known shared frailty model. This model provides a suitable theoretical basis for generating dependence between the components’ failure times in the dependent competing risks model. It is known that the dependence effect in this scenario influences the estimates of the model parameters. Hence, under the assumption that the cause-specific intensities follow a PLP, we propose a frailty-induced dependence approach to incorporate the dependence among the cause-specific recurrent processes. Moreover, the misspecification of the frailty distribution may lead to errors when estimating the parameters of interest. Because of this, we considered a Bayesian nonparametric approach to model the frailty density in order to offer more flexibility and to provide consistent estimates for the PLP model, as well as insights about heterogeneity among the systems. Both simulation studies and real case studies are provided to illustrate the proposed approaches and demonstrate their validity.
\end{abstract}


\section*{ACRONYMS AND ABBREVIATIONS}
\vspace{-0.35cm}
\begin{table}[ht]
{\normalsize
\begin{tabular}{l l } 
Acronyms &  \\
PLP & Power law process.\\
HPP & Homogeneous Poisson process.\\
NHPP & Nonhomogeneous Poisson process.\\
DPM & Dirichlet process mixture.\\
HMC & Hamiltonian Monte Carlo.
\end{tabular}}
\end{table}

\section*{NOTATION}
\vspace{-0.25cm}
\begin{table}[ht]
{\normalsize
\begin{tabular}{l l } 

\end{tabular}}
\end{table}

\begin{table}[ht]
{\normalsize
\begin{tabular}{l l } 
$\hat\beta_{j}^{Bayes}$ & Bayes Estimator of $\beta_j$.\\
$\hat\alpha_{j}^{Bayes}$ & Bayes Estimator of $\alpha_j$.\\
$\lambda(\cdot)$ & Intensity function.\\
$N(\cdot)$ & Univariate counting process.\\
$\alpha \ \hbox{and} \ \beta$ & PLP parameters.\\
$\Lambda(\cdot)$ & Mean function.\\
$Z$ & Frailty variable.\\
\end{tabular}}
\end{table}

\IEEEpeerreviewmaketitle

\vspace{0.5cm}

\section{Introduction}

\vspace{0.3cm}

\IEEEPARstart{S}{tudying} recurrent event data is important in many areas such as engineering, social and political sciences and in the public health setting. In all these fields of study, the event of interest occurs on a recurring basis. For example, failure of a mechanical or electrical component may occur more than once; the recurrence of bugs over time in a software system that is under development; successive tumors in cancer studies; myocardial infarction and epileptic seizure in patients, to name but a few. 

In particular, in reliability analysis, interest is usually centered on failure data from complex repairable systems \cite{ascher1984repairable}. Monitoring the status of a repairable system leads to a recurrent events framework, where events correspond to failures of a system. A system is defined as repairable when it receives any corrective measure (other than replacing the whole system) in order to restore its components when they have failed and can be returned to the satisfactory operation state where it is able to perform all its functions. On the other hand, a nonrepairable system is a system that is discarded when the first failure occurs \cite{rigdon2000statistical}. However, we will just focus on repairable system case.

The primary challenge when modeling repairable systems data is how to account for the effect of a repair action performed immediately after a failure. In general, one assumes that repair actions are instantaneous and repair time is negligible. The most explored assumptions are either minimal repair and perfect repair. In the former, it is supposed that the repair action, after a failure, returns the system to the exact condition it was immediately before it failed. In the latter, the repair action leaves the system as if it were new. 
In the engineering literature, these types of repair or corrective 
maintenance are usually called: ABAO and AGAN \cite{barlow1960optimum, aven1983optimal, aven2000general,finkelstein2004minimal, mazzuchi1996bayesian}.
More sophisticated models which account for repair action that leave the system somewhere between the ABAO and AGAN conditions are possible, although they will not be considered here; see for instance, 
\cite{doyen2006imperfect}.


Statistical modeling of the occurrence of failures is done using point processes, particularly, as we will see later, counting processes. In this framework, the model is completely characterized by its failure intensity function. The failure history of a repairable system, under a minimal repair strategy, is usually modeled according to a NHPP. In the repairable system literature, one of the most important and well-known parametric forms for the NHPP model is the PLP. The PLP process is convenient because it is easy to implement, it is flexible and the parameters have good interpretations. Regarding classical inference for the PLP, see, for instance, 
\cite{ascher1984repairable} or 
\cite{rigdon2000statistical}.
Bayesian inference has been considered 
among others by 
\cite{bar1992bayesian},  \cite{guida1989bayes},  \cite{pievatolo2004bayesian} and  \cite{ruggeri2006reliability}.

Additionally, in this work, we emphasize an alternative specification of the PLP, which is obtained by using a simple operational definition of its parameters, making them orthogonal to each other. This formulation is considered by \cite{de2012bayesian} motivated by ideas from 
\cite{guida1989bayes} and \cite{sen2002bayesian}. The former authors show that this reparametrization leads to some advantageous results such as orthogonality among parameters, the likelihood function becomes proportional to a product of gamma densities and the expected Fisher information matrix is diagonal. The model we discuss here is based on such reparametrization because it results in mathematical and computational simplifications for our research.

In reliability theory, the most common system configurations are series systems, parallel systems, and series-parallel systems. In a series system, components are connected in series, in such a way that the failure of a single component results in system failure. The same setting may be expressed in an alternative way by a repairable system in which components can perform different operations, and thus be subject to different types of failures. Traditionally, models with this characteristic are known as competing risks. In complex systems, such as supercomputers, aircraft generators, industrial plants, jet engines, and cars,  the presence of multiple types (or causes) of failure is common. From an economic perspective, such systems are commonly repaired rather than replacing the system with a new one after failure. Thus, this model can also be called a repairable competing risks system. As we pointed out already, commonly used methodologies for analyzing multi-type recurrent event data are based on multivariate counting processes and cause-specific intensity functions \cite{andersen2012statistical, cook2007statistical}. 

It is worth noticing that the existing literature on competing risks in reliability is extensive and focuses particularly on analysis for nonrepairable systems, e.g., \cite{crowder2001classical},  \cite{lawless2011statistical}, \cite{crowder1994statistical} to cite a few.
On the other hand, a number of authors have considered modeling competing risks in a repairable systems framework. For example, some authors have mainly been interested in questions concerning maintenance analysis \cite{langseth2006competing, lindqvist2006statistical, doyen2006imperfect}. Others have highlighted the relevance of failure analysis of the components of the system based on cause-specific intensity function, such as \cite{liu2010accelerated, fu2014objective, somboonsavatdee2015parametric, somboonsavatdee2015statistical}, \cite{ pollo2018objective, almeidastatistical}. 

In the field of reliability engineering, much of the current literature on competing risks pays particular attention to the hypothesis that the components' (causes) failures are independent from each other \cite{liu2010accelerated, hong2010field, meeker2014statistical,  somboonsavatdee2015statistical}. However, this assumption is restrictive in some real situations because there are many ways of dependence between components. We can call this case dependent competing risks. Moreover, it is important to point out that neglecting existing dependence can lead to estimation errors and bad predictions of system behavior \cite{zhang2015optimal}. A seminal study in this area is the work of \cite{moeschberger1974life}. \cite{wu2017statistical} and \cite{zhang2017system} give an extensive discussion on copula theory in order to model the dependence between components (competing failure modes) in particular settings. \cite{zhang2015optimal} discuss an optimal maintenance planning for dependent competing risks systems. \cite{liu2012planning} mentions a particular situation where the components within a system are physically, logically, or functionally connected, as an example of dependent failure causes. It means that the condition of a component influences or induces the failure of other components and vice-versa. This author works with the dependence framework based on a gamma frailty model. An interesting perspective has been explored by \cite{lindley1986multivariate}, who argue that dependence can be induced by the environment the system is subjected to, i.e., the situation where the components of the system (or cluster) share the same environmental stress. Along the same lines, \cite{somboonsavatdee2015parametric} assert that in a repairable systems context such clustering arises naturally across the recurrent failures of a system. The approach proposed by \cite{somboonsavatdee2015parametric} for modeling dependence is also based on frailty.

These examples demonstrate the importance of the theme and, therefore, the need to develop new analysis methodologies. However, very few articles address the dependence, particularly in the setting of recurrent competing risks in repairable systems with PLP. Based on these reasons, we propose a shared frailty \cite{wienke2003frailty, hougaard2012analysis} model using a (multivariate) counting process framework whose intensity function is that of reparametrized PLP. Specifically,  the intensity is multiplied by a frailty (or random effect) term, which follows a suitable distribution for a positive random variable. This model provides a suitable theoretical basis for generating dependence between the components’ failure times. In other words, the components belonging to a cluster (or system) share a common factor (frailty term), which generates such dependence. The assignment of a probability distribution to frailty plays an important role in the analysis of models with random effects. However, in order to avoid making incorrect model specifications when there is uncertainty about some inherent characteristics of a distribution (e.g., multimodality, skewness, and heavy tails)  \cite{walker1997hierarchical}, we propose a nonparametric approach to model the frailty density (density estimation) \cite{ferguson1973bayesian, ferguson1974prior, muller2004nonparametric}. Our approach to these problems is fully Bayesian and based on both MCMC methods (for frailty) and closed-form Bayesian estimators (for PLP parameters) for estimation. 

The main contributions of the proposed research include: (i) Our modeling of the dependence effect on multi-component systems, based on multivariate recurrent  processes and the shared frailty model, is advantageous because we can perform an individual posterior analysis of the quantities of interest, i.e., we estimate the interest parameters of the PLP (our main focus) separately from nuisance parameter of frailty distribution (variance). Regarding PLP parameters, we consider noninformative priors so that the posterior distributions are proper. With respect to frailty, our proposal avoids making incorrect specifications of the frailty distribution when there is uncertainty about some inherent characteristics of distribution. In this case, we use nonparametric Bayesian inference. Besides, a particular novelty is our hybrid MCMC algorithm for computing the posterior estimates with respect to the frailty distribution, see \cite{almeidastatistical}.

In this paper, the main objective is to study certain aspects of modeling failure time data of repairable systems under a competing risks framework. We propose more efficient Bayesian methods for estimating the model parameters. Thus, we can list some specific objectives: 
\begin{itemize}
    \item to consider an orthogonal parametrization for the PLP model parameters such that the likelihood function becomes proportional to a product of gamma densities and the expected Fisher information matrix is diagonal;
    \item to propose a frailty-induced dependence approach to incorporate the dependence among the cause-specific recurrent processes. Besides, to consider a nonparametric approach to model the frailty density using a DPM prior. Additionally, to propose a hybrid MCMC sampler algorithm composed by HMC and Gibbs sampling to compute the posterior estimates with respect to the frailty distribution. Regarding PLP parameters, to propose a class of noninformative priors whose resulting posterior distributions are proper and to obtain closed-form Bayesian estimators.
\end{itemize}

The remaining part of the paper proceed as follows: In Section \ref{secII}, fundamental groundwork in terms of repairable systems, dependent competing risks and shared frailty model is presented. Section \ref{secIII}, presents the modeling of the problem in a point of view of multiple repairable systems under dependent causes of failure. The Bayesian analysis is developed in Section \ref{secIV}, with a discussion on the choice of priors distributions for the proposed model and the computation of posterior distributions. In Section \ref{secV}, an extensive simulation study is described in order to evaluate the efficiency of the proposed Bayesian estimators, and Section \ref{secVI} uses them to analyse a real data set that comprises the failure history for a fleet of cars under warranty. Section \ref{secVII} concludes the paper with final remarks.

\section{Background} \label{secII}

\subsection{Multiple repairable systems}

In this section, we present a brief overview to analyze data from multiple repairable systems, but we refer the reader to \cite{rigdon2000statistical} and \cite{de2012bayesian} for details and proofs. Here, we highlight just two important assumptions in this context. The first is to assume that all systems are identical or different. The second is to assume that all systems have the same truncation time $\tau$  or, otherwise, have different truncations at $\tau_j$. However, for the sake of simplicity and brevity of exposition, we assumed the observation lengths, $\tau$, for each system to be equal.  Moreover, in this paper, we assume all systems to be identical, i.e., the systems are specified as $m$ independent realizations of the same process, with intensity function $\lambda$.

If the multivariate counting processes $N_1(t), \dots, N_m(t)$ are all observed at the same time $\tau$, the NHPP resulting from the superposition of NHPPs is given by $N(t)=\sum_{j=1}^m N_j(t)$ and has an intensity function given by $\lambda(t) = m\lambda(t)$; e.g., overlapping realizations of a PLP. Therefore, inferences in models proposed for this framework can be made through the following likelihood function
\begin{equation*}
    L(\lambda) = \left( \prod_{j=1}^m \prod_{i=1}^{n_j}  \lambda(t_{ji})   \right) \exp \left( - \sum_{j=1}^m \int_{0}^{\tau} \lambda(s)ds \right).
\end{equation*}

\subsection{Unobserved heterogeneity between multiple systems}
The $m$ systems are considered to be identical, and therefore have the same intensity function and thus we would have a random sample of systems. On the other hand, this assumption may not be true. That is, in many real-world reliability applications there may be some heterogeneity between "apparently identical" repairable systems. In this case, it is necessary to propose a statistical model capable of capturing this heterogeneity. \cite{cha2014some, asfaw2015unobserved, slimacek2016nonhomogeneous} and \cite{slimacek2017nonhomogeneous} discuss frailty models for modeling and analyzing repairable systems data with unobserved heterogeneity.

\subsection{The minimal repair model}
Before turning to formal definitions, we provide an intuitive and real example. We said earlier that a minimal repair policy is enough to make the system operational again. For example, if the water pump fails on a car, the minimal repair consists only of repairing or replacing the water pump. As we said before, the purpose is to bring the car (system) back to operation as soon as possible. From an economic perspective, complex systems are commonly repaired rather than replacing the system with a new one after failure.

Recalling that NHPP is completely specified by its intensity function, then when parametric models are adopted for the intensity function of an NHPP, we are interested in making inferences about the parameters of this function. In addition, one knows that the NHPP forms a class of models that naturally applies to a “minimal repair”, i.e., the repair brings the system back into the same state it was in just prior to the failure. One of the most important and used functional forms is the PLP. 

The parametric form for the PLP intensity is given by
\begin{eqnarray}\label{equanhpp2}
\lambda (t)=(\beta / \mu) (t / \mu )^{\beta -1},
\end{eqnarray}
where $\mu, \beta>0$. Its mean function is 
\begin{eqnarray}\label{equanhpp3}
\Lambda(t) = \mathbb{E}[N(t)]=\int_{0}^{t} \lambda (s) ds =(t / \mu )^{\beta}.
\end{eqnarray}

The scale parameter $\mu$ is the time for which we expect to observe a single event, while $\beta$ is the elasticity of the mean number of events with respect to time \cite{de2012bayesian}.

Since (\ref{equanhpp2}) increases (decreases) in $t$
for $\beta > 1$ ($\beta < 1$), 
the PLP can accommodate both systems that deteriorate 
or improve over time. Of course, when $\beta =1$, the 
intensity (\ref{equanhpp2}) is constant and hence the PLP 
becomes an HPP.

Under minimal repair, the failure history of a repairable system is modeled as an NHPP. As mentioned above, 
the PLP (\ref{equanhpp2}) provides a flexible parametric form for the intensity of the process. 
Under the {\em time truncation} design, i.e.\ when failure data is 
collected up to time $T$, the likelihood becomes
\begin{eqnarray}\label{jointpdf}
L(\beta,\mu \mid n, \textbf{t})=\frac{\beta^n}{\mu^{n\beta}} \left(\prod_{i=1}^{n} t_i \right)^{\beta -1}\hspace{-0.25cm}\exp\left[ - \left( \frac{T}{\mu}\right)^{\beta} \right],
\end{eqnarray}
where we assume that $n \geq 1$ 
failures at times $t_1<t_2<\dotsc<t_n<T$ were observed, $i=1, \dots, n$ \cite{rigdon2000statistical}. The MLEs of the parameters are given by
\begin{equation}\label{mleclassic}
    \hat{\beta} = n/\sum_{i=1}^{n} \log (T / t_i) \ \ \hbox{ and } \ \ \ \hat{\mu} = T / n^{ 1/ \hat{\beta} }.
\end{equation} 

\subsection{Reparametrized PLP}

\cite{de2012bayesian} suggest reparametrizing the model (\ref{equanhpp2})
in terms of $\beta$ and $\alpha$, where the latter is given by
\begin{eqnarray}\label{reparam}
\alpha = \mathbb{E}[N(T)]=(T/\mu)^{\beta},
\end{eqnarray}
so that the likelihood (\ref{jointpdf}) becomes
\begin{equation}\label{eqverplp}
\begin{aligned}
L(\beta, \alpha| n, \textbf{t})&=c\left(\beta^n e^{-n\beta/\hat{\beta}}\right)\left(\alpha^n e^{-\alpha}\right)  \\ &
\propto \gamma(\beta|n+1, n/\hat{\beta})\gamma(\alpha|n+1, 1)\,, 
\end{aligned}
\end{equation}
where $c=\prod_{i=1}^{n} t^{-1}_{i}$, $\Hat{\beta} = n/\sum_{i=1}^{n} \log (T / t_i) $ is the MLE of $\beta$ and $\gamma(x|a, b)=b^ax^{a-1}e^{-bx} / \Gamma(a)$ is the PDF of the gamma distribution with shape and scale parameters $a$ and $b$, respectively. It is important to point out that $\beta$ and $\alpha$ are orthogonal parameters. For the advantages of having orthogonal 
parameters, see \cite{cox1987parameter}.

\subsection{Competing risks}
\begin{figure}[!h]
    \centering
    \includegraphics[width=6cm]{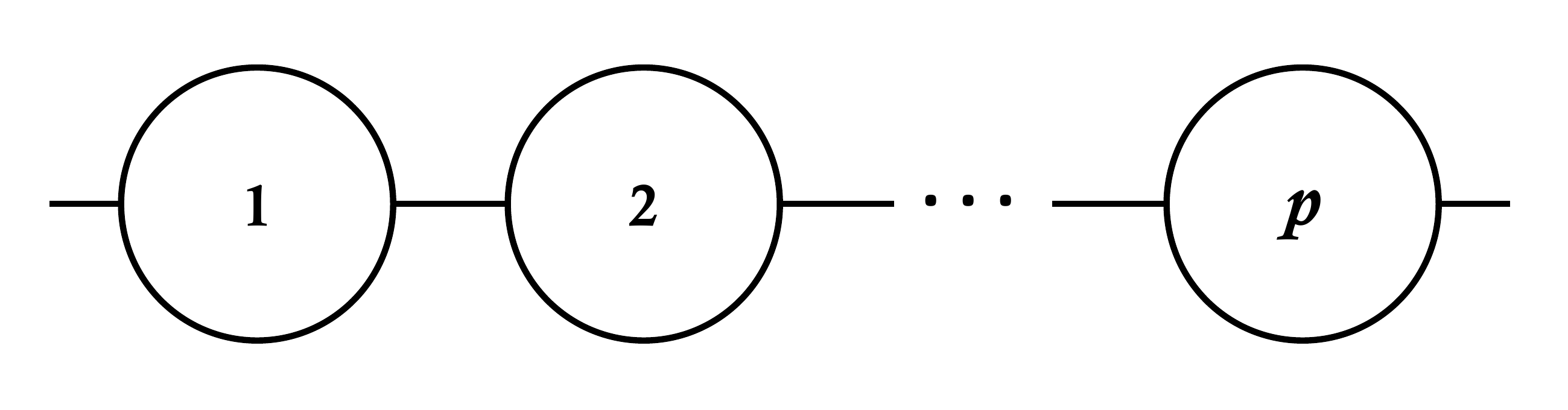}
    \caption{Diagram for a competing risks system (i.e., series system) with $p$ risks (components).} \label{figure_cp}
\end{figure}

In reliability theory, the most common system configurations are the series systems, parallel systems, and series-parallel systems. 
In a series system, components are connected in such a way that the failure of a single component results in system failure. Such a system is depicted in Figure \ref{figure_cp}. 
A series system is also referred to as a competing risks system since the failure of a system can be classified as one of the $p$ possible risks (components) that compete for the failure of the system. 
In general, the observations of a competing risks model consist of the pair ($t$, $\delta$), where $t \geq 0$ represents the time of failure and $\delta$ is the indicator of the component which failed. An example follows to illustrate the failure history data for this kind of framework. 

\begin{figure}[!h]
    \centering
    \includegraphics[width=8.5cm]{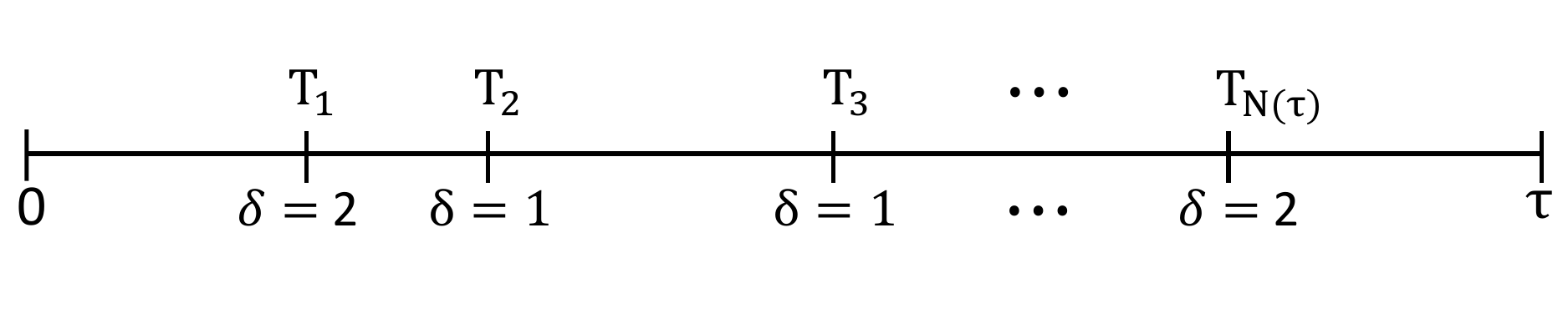}
    \caption{Observable quantities from failure history of a repairable competing risks system with two recurrent causes of failure.} \label{figure_retaRC}
\end{figure}

{\em example:} Suppose a repairable system, and let $0<T_1<T_2<T_3<\dots<T_{N(\tau)}<\tau$ be the failure times of the system observed until a pre-fixed time $\tau$. Moreover, there are two ($p = 2$) recurrent causes of failure, and at the {\em i}-th failure time $T_i$, we also observe $\delta \in \{1, 2 \}$, which is the cause of the failure related to the {\em i}-th failure (see Figure \ref{figure_retaRC}). 

Basically, we could say that in most of the literature, there are two main approaches when  analyzing failure times with competing risks: independent and dependent competing failure modes. For reliability models under competing risks, most research has been carried out considering statistical independence of component failure \cite{crowder1994statistical, hoyland2009system, somboonsavatdee2015statistical, todinov2015reliability, wu2017two}. Thus, one assumes that independent risks are equivalent to independent causes of failure.
However, in some particular contexts (for instance, the existence of interactions between components in complex systems), the assumption of independent risks may lead to seriously misleading conclusions. To overcome this issue, some important and general approaches have been presented in the literature for modeling dependent competing risks data \cite{dijoux2009alert, zhang2015optimal, zhang2017system}.

Considering Daniel Bernoulli's attempt in the 18th century to separate the risk of dying due to smallpox from other causes \cite{bernoulli1760essai, bradley1971smallpox}, the competing risks methodology has disseminated through various fields of science such as demography, statistics, actuarial sciences, medicine and reliability analysis. Therefore, one knows that both the theory and application of competing risks is too broad to cite here, but for an overview of the basic foundations, please see \cite{pintilie2006competing, crowder2001classical, crowder2012multivariate}. For repairable systems failing due to competing risks, we refer the reader to \cite{langseth2006competing, doyen2006imperfect, somboonsavatdee2015statistical}. Particularly, this article responds directly to the application in repairable systems under a recurrent data structure based on stochastic processes, which is the most natural way to describe the recurrence of multiple event types that occur over time. 

\subsection{Recurrent competing risks model for a single repairable system}\label{oneSistRCsection}

The assumption of the repairable system under examination is that the components can perform different operations, and thus be subject to different types of failures. Hence, in our model there are $K$ causes of failure. If $n$ failures have been observed in $(0, T]$, then we observe the data $(t_1, \delta_1), \dots, (t_n, \delta_n)$, where $0 <t_1 <\dots < t_n < T$ are the system failure times and $\delta(t_i)=\delta_i = q $ represents the $q$-th associated failure cause with $i$-th failure time, $i=1,\dots,n$ and $q=1,\dots,K$. 

One can introduce a counting process $ \{ N_q(t); t\geq0 \}$ whose behavior is associated with the cause-specific intensity function
\begin{footnotesize}
\begin{equation}
\begin{aligned}
\lambda_q (t) = \lim_{\Delta t \rightarrow 0} \frac{P(\delta (t) = q, N( t + \Delta t] - N(t) = 1 \mid N(s), 0 \leq s \leq t)}{\Delta t}.
\end{aligned}
\end{equation}
\end{footnotesize}

Let $N_q$ be the cumulative number of observed failures for the $q$-th cause of failure and $N(t)=\sum_{q=1}^{K} N_q(t)$ be the cumulative number of failures of the system. Thus, $N (t)$ is a superposition of NHPPs and its intensity function is given by
\begin{equation}
\begin{aligned}
\lambda (t ) & = \lim_{\Delta t \rightarrow 0} \frac{P( N( t + \Delta t] - N(t) = 1 \mid N(s), 0 \leq s \leq t)}{\Delta t} \\
&= \sum_{q=1}^{K} \lambda_q (t).
\end{aligned}
\end{equation}

The cause-specific and the system cumulative intensities are, respectively,
    \begin{equation} 
    \Lambda_q ( t )=\int_{0}^{ t } \lambda_q(u ) du \ \ \hbox{ and } \ \ \Lambda ( t )= \sum_{q=1}^{K} \Lambda_q( t ).
    \end{equation}

Under minimal repair, the failure history of a repairable system is modeled as an NHPP. We give special attention to functional form for the cause-specific intensity according to the PLP, as follow
\begin{equation}\label{plpCR}
    \lambda_q (t )= \frac{\beta_q}{\psi_q} \left( \frac{t}{\psi_q} \right)^{\beta_q - 1},
\end{equation}
with $t\geq 0$, $\psi_q>0$, $\beta_q >0$ and for $q=1, \dots, K$. The model is quite flexible because it can accommodate both decay ($\beta_q<1$) and growth ($\beta_q>1$) in reliability.
The corresponding mean function considering time-truncated scenario (with fixed time $T$) is
\begin{equation}\label{meanPLP}
    \mathbb{E}[N_q(T)] = \Lambda_q(T) = \left( \frac{T}{\psi_q} \right)^{\beta_q}.
\end{equation}
If we reparametrize (\ref{plpCR}) in terms of $\beta_q$ and $\alpha_q$, where the latter is given by
\begin{equation}\label{plp_reparametrized}
    \alpha_q = \left(  \frac{T}{\psi_q} \right)^{\beta_q},
\end{equation}
one obtains the following advantageous likelihood function
\begin{equation}
\begin{aligned}\label{verossim1}
L(\boldsymbol\theta|\textbf{t},\boldsymbol\delta) &= \left\lbrace \prod_{ i=1 }^{ n } \prod_{q=1}^{K}  \left[ \beta_q \alpha_q t_i^{\beta_q - 1} T^{-\beta_q} \right]^{\mathbb{I} (\delta_{i}=q)} \right\rbrace \\
& \quad \times \exp \left\lbrace \sum_{q=1}^K  \alpha_q \right\rbrace \\
&\propto \prod_{q=1}^{K}\gamma(\beta_q|n_{q}+1, n_{q}/\hat{\beta_q}) \prod_{q=1}^{K}\gamma(\alpha_q|n_{q}+1, 1),
\end{aligned}
\end{equation}
where $n = \sum_{q=1}^{K} n_q$; $n_q = \sum_{i=1}^{n} \mathbb{I} ( \delta_i = q) $; $\boldsymbol\theta=(\boldsymbol{\beta, \alpha})$ with $\boldsymbol{\beta}=(\beta_1, \dots, \beta_K)$ and $\boldsymbol{\alpha}=(\alpha_1, \dots, \alpha_K)$; $\hat{\beta}_q = n_q / \sum_{i=1}^{n_q} \log (T / t_i) = n_q / \sum_{i=1}^{n} \log (T / t_i)\mathbb{I} ( \delta_i = q)$.

\subsection{Frailty model}\label{Frasection}

Frailty models are generalizations of the well-known Cox model \cite{cox1972regression}, introduced by \cite{vaupel1979impact}. Over the past decades, most research in frailty has emphasized the analysis of medical and reliability data that present heterogeneity, which cannot be adequately explained by the Cox model. To be more precise, it can be said that, the frailty term is a random effect that acts multiplicatively on the hazard function of the Cox model. This random effect could represent misspecified or omitted covariates (unknown or unmeasured effects). Thus, one can say that such a term (frailty) is an unobservable (latent) quantity. In addition, the frailty methodology is very effective to account for dependency in event times that result from unknown sources of heterogeneity. For more details on general frailty theory, see  \cite{hougaard2012analysis, hanagal2011modeling}.

Considering recurrent event data,  several approaches of the Cox model with a frailty factor have been discussed in the literature \cite{tomazella2003modelagem, bernardo2011objective}. Additional results on frailty in the counting process context are given in \cite{aalen2008survival} and \cite{andersen2012statistical}. In the reliability field, the frailty model is commonly used to model heterogeneous repairable systems \cite{cha2014some, asfaw2015unobserved}. Such heterogeneity is generated because some units have a higher (or lower) event rate than other units due to unobserved or unknown effects (e.g., instability of production processes, environmental factors, etc.). On the other hand,  \cite{somboonsavatdee2015parametric} present a classical inference for repairable systems under dependent competing risks where the frailty is considered to model the dependence between the components arranged in series.

The many approaches differ in the modeling of the baseline hazard or in the distribution of the frailty. There is a vast amount of published studies describing fully parametric approaches. Regarding the probability distribution that should be assigned to the frailty term (random effect), in general, it follows a distribution appropriate for a positive random variable. Parametric frailty models are standard in the literature \cite{aalen2008survival, hougaard2012analysis} and the so-called gamma frailty model, in which the unobserved effects are assumed to be gamma distributed, is probably the most popular choice. Various frailty distributions are presented in \cite{hougaard2012analysis} and the references therein such as the gamma, inverse Gaussian, log-normal or the positive stable frailty. Other distributions include the power variance frailty \cite{aalen2008survival} and the threshold frailty \cite{lindley1986multivariate}. 

Extensive research has been carried out on frailty distributions, as cited above, and it is well known that, generally, such distributions are primarily used by mathematical convenience. Furthermore, in general, such distributions do not encompass a range of possible features including skewness and multimodality. Furthermore, because the frailty variable is an unobservable quantity, it cannot be tested to verify whether or not it satisfies the distributional assumption \cite{ferreira2001simulation}. It is known that the misspecification of this distribution can lead to several types of errors, including, for example, poor parameter estimates \cite{walker1997hierarchical}. A more flexible and robust approach would be to estimate such a density using the nonparametric Bayesian methodology.  In this work, this solution will be explored.

Finally, a suitable choice of the distribution of unobserved effects can provide interesting general results, but generally the main quantity of interest is the variance of the unobserved effects. Usually, a significant variance may indicate high dependence \cite{wienke2003frailty,somboonsavatdee2015parametric}. 

\subsection{Shared frailty}
In order to emphasize the subject matter of the repairable systems under dependent competing risks, we introduce the shared frailty model using the multivariate counting processes framework based on cause-specific intensity functions.

The referred dependency between competing risks may be modeled through a frailty variable, say $Z$, in such a way that, when the frailty is shared among several units in a cluster, it leads to dependence among the event times of the units \cite{wienke2003frailty, tomazella2003modelagem, hougaard2012analysis}. Suppose that $m$ clusters (or systems) are under observation, where each cluster is composed by $K$ units (or components). The intensity function of the $j$-th cluster ($j = 1, \dots, m$) of a shared frailty model is that of the Cox model multiplied by a frailty term $Z_j$ (multiplicative random effect model). More specifically, for each individual counting process, $\{ N_{j}(t): t \geq 0 \}$, their intensity function, conditionally on the frailty $Z_j$, is given by 
\begin{equation}\label{funcintensZ}
\lambda_j ( t \mid Z_j) = Z_j \lambda(t),
\end{equation}
where $\lambda(t)$ is the basic intensity function and  $j = 1, 2, \dots, m$. The intensity function (\ref{funcintensZ}) describes the recurrent failure process on the $j$-th cluster and the intensity associated to the $q$-th component from the $j$-th cluster is defined as
\begin{equation}\label{funcintensZunit}
\lambda_{jq} (t \mid Z_j) = Z_j \lambda_q(t),
\end{equation}
where $\lambda_q(t)$ is the basic intensity function from the $q$-th component (cause-specific intensity function), $q = 1, 2, \dots, K$. Note that intensities (\ref{funcintensZ}) and (\ref{funcintensZunit}) follow the relation $\lambda(t) = \sum_{q=1}^{K} \lambda_q(t)$ \cite{andersen2012statistical}. Henceforth, we will omit the subscript $j$ from $\lambda_{jq}$ in (\ref{funcintensZunit}) since we are assuming that the systems are identical, therefore $ \lambda_{q} (t \mid Z_j ) = Z_j \lambda_{q}^{'}(t)$. Note also that $\lambda_{q}^{'}(t)$ is referred to as the (basic) intensity function for type $q$ events (e.g., PLP). Let $\textbf{Z} = (Z_1, \dots , Z_m)$ denote the vector of $Z_j$s, which we assume arises from density  $f_{Z}(\cdot)$, where each $Z_j$ is iid. These are typically parametrized so that both  $E(\boldsymbol{Z})$ and $Var(\boldsymbol{Z})$ are finite, for $j = 1,2,... ,m$. It is worth pointing out that the $Z_j$s are assumed to be stochastically independent of the failure process $\lambda_{q}^{'}(t)$ \cite{andersen2012statistical, somboonsavatdee2015parametric}.

The term of frailty in (\ref{funcintensZ}) aims to control the unobserved heterogeneity among systems. If we consider the situation  where the dataset is divided into clusters (multiple units in a cluster), this term evaluates the dependence between the units that share the frailty {\em $Z_j$}. Thus, units from heterogeneous populations can be considered independent and homogeneous, conditionally to the terms of frailty ($Z_j$s) attributed to the units or cluster of units.

The evaluation of the influence of unobserved heterogeneity in this type of data is made on the basis of the variability of the frailty distribution. In addition, it is worth pointing out that higher values of $Var(Z)$ mean greater heterogeneity among units and more dependence between the event times for the same unit. In general, in the literature, it is common to specify a distribution for the frailty variable with mean 1 and variance, say $Var(Z)=\eta$, in order to obtain two main advantages: (1) the model parameters become identifiable, and (2) it is possible to obtain an easily understandable interpretation of the model, because, as previously argued, $\eta$ acts as a dependence parameter, meaning that, if the frailty variance is zero, it implies that we have independence between the event times in the clusters (since it is assumed that the mean is 1).

We emphasize here that the main focus of SECtion of the present work will be to analyze data from multiple repairable systems under the presence of dependent competing risks. Thus, to estimate the model parameters considering shared frailty. In this sense, naturally, we can consider the dependence between the components as a nuisance parameter via frailty.

\section{The model} \label{secIII}

As mentioned before, the nonparametric frailty distribution takes into account a flexible class of distributions. In particular, we used the DPM model to describe the frailty distribution due to its flexibility in modeling unknown distributions. Many approaches on nonparametric Bayesian models have been explored in the literature related to reliability, for instance, \cite{salinas2002bayesian} give a comprehensive theoretical exposition on Bayesian nonparametric estimation for survival functions arising from observed failures of a competing risks model (or a series system). \cite{li2014bayesian} provide a flexible Bayesian nonparametric framework to modeling recurrent events in a repairable system to test the minimal repair assumption. Bayesian nonparametric inference for NHPPs is considered by \cite{kuo1997bayesian}, who employed several classes of nonparametric priors. As mentioned above, the idea of our approach is to apply a Bayesian nonparametric prior (i.e., DPM prior) to modeling uncertainty in the distribution of shared frailty. Although this model has infinite parameters, due to the infinite mixture model, it is a flexible mixture, parsimonious and simple to sample. We chose the stick-breaking representation of the DP prior \cite{sethuraman1994constructive}, because of a simple implementation to build the algorithm. To obtain the posterior distribution, we created a hybrid MCMC algorithm \cite{kalli2011slice}, using the Gibbs sampler \cite{smith1993bayesian} and the HMC method \cite{neal2011mcmc}. It is important to point out that no studies have been found which explore the use of DPM for frailty density in the context of multiple repairable systems under the action of dependent competing risks. 

This research highlights the importance of modeling the dependence structure among competing causes of failure by using a more flexible distribution for unknown frailty density in order to provide good estimates of the model parameters. As stated before, our primary inference goal is to estimate PLP parameters. To this end, firstly, we model the dependence effect with shared frailty, and secondly, we consider the frailty distribution nonparametrically using a DPM. Regarding frailty,  the advantage is that one obtains more flexibility at the level of density estimation and providing insights in terms of heterogeneity among systems.

\subsection{Multiple repairable systems subject to multiple causes of failure} \label{multSystRCsection}

Here, we highlight the used notations in the multivariate counting process context. Hereafter, random variables are denoted by capital letters (e.g., $Z_j$, $N_{jq}$), while their realizations are denoted by the lowercase (e.g., $z_j$, $n_{jq}$).

Consider a sample of $m$ identical systems in which each system is under the action of $K$ different types of recurrent causes of failure. Let $N_{jq}(t)=\sum_{i=1}^{n_j} \mathbb{I}(\delta_{ji}=q)$  be the cumulative number of type $q$ failures occurring over the interval $[0,t]$ for the $j$-th system ($j = 1, \dots , m$; $q=1, \dots, K$ and $i=1, 2, \dots, n_{j}$), where $\{ N_{jq}(t): t \geq 0 \}$ is a counting process. Note that $N_{j\bullet} (t) = \sum_{q=1}^{K} N_{jq} (t)$ represents the cumulative number of failures of system $j$ taking into account all failures arising from all components from the $j$-th system. Let $N_{\bullet q} (t) = \sum_{j=1}^{m} N_{jq} (t)$ denote the number of failures of cause $q$ for all systems. 

Suppose that each system is under observation for all types of events over the same period of time, i.e., $[0, T]$. Thus, let $t_{ji}$, $i=1, 2, \dots, n_{j}$, be the observed failure times for system $j$, satisfying $ 0 < t_{j1} < t_{j2} < \dots < t_{jn_j} < T $. Besides, denote that $\delta_{ji}=q$ is the failure mode (or component) that caused the system failure. Hence, the observed data is $D_j = \{ (t_{ji}, \delta_{ji}=q), i=1, 2, \dots, n_{j}; q=1, \dots, K \}$. The complete data is given by $\boldsymbol{D}=(D_1, \dots, D_m)$.

As mentioned earlier, our focus is mainly on the component level failure process which conforms to a PLP, therefore the cause-specific intensity function that governs the counting process $N_{\bullet q}(T)$, taking into account a orthogonal parametrization in terms of $\alpha_q$ and $\beta_q$, is defined as
\begin{equation}\label{plpCRrepar}
\lambda_{q} ( \boldsymbol{t} ) = \beta_q \alpha_q t^{\beta_q -1} T^{-\beta_q}, \ q=1, \dots, K,
\end{equation}
where $\alpha_q$ is the mean function given by
\begin{equation}\label{meanplpCRrepar}
\alpha_q = \mathbb{E} \left[ N_{\bullet q} (T) \right] = \Lambda_{q} ( T ) = \int_{0}^{T} \lambda_{q} (s) ds.
\end{equation}

\subsection{The shared frailty model for the PLP}\label{Fra2section}

It is worth pointing out that the main quantity of interest in the shared frailty methodology adopted here is the variance of the frailty (although it is considered as a nuisance parameter, because our major interest is to estimate the PLP parameters). This parameter should be estimated using information of multiple systems. \cite{somboonsavatdee2015statistical} state that in the single system setting there are limitations. Therefore, our approach requires multiple systems as presented so far.

We specify the model (\ref{funcintensZunit}) in terms of (\ref{plpCRrepar}) in order to 
present the likelihood function with a special form. To achieve this purpose, suppose a minimal repair is undertaken at each failure, thus the NHPP is the model of choice. Specifically, the failures from each component follow an NHPP, with PLP intensity function given in (\ref{plpCRrepar}). Furthermore, let us consider that a realization $z_j\sim f_Z$ acts on all the cause-specific intensities (\ref{funcintensZunit}) belonging to the  $j$-th system. Thus, conditioning on the frailty term, the model is expressed as
\begin{equation}\label{plpCRZmulti}
\lambda_{q} ( \boldsymbol{t} | Z_j) = Z_j \beta_q \alpha_q t_{ji}^{\beta_q -1} T^{-\beta_q}   
\end{equation}
and the mean function is given by
\begin{equation}\label{meanplpCRZmulti}
\Lambda_{q} ( T | Z_j) = Z_j \alpha_q.
\end{equation}

It is important to point out that, hereafter, our analysis relies on the constraint $ \bar{\textbf{Z}}=\frac{1}{m} \sum_{j=1}^{m} Z_j = 1 $.

\subsection{Likelihood function}\label{likelimodelsection}
To simplify notation in this section, we will drop the subscript $\bullet$ and refer to $n_{j\bullet}$ and $n_{\bullet q}$ as $n_{j}$ and $n_{q}$, respectively. The likelihood contribution from the $j$-th system based on (\ref{plpCRZmulti}) is given by

\begin{eqnarray}\label{verossim2}
\begin{aligned}
L_j(\boldsymbol\theta, Z_j| D_j ) &= \left[ \prod_{i=1}^{n_j}\prod_{q=1}^{K} [ \lambda_{q} ( t_{ji} | Z_j) ]^{\mathbb{I}(\delta_{ji}=q)} \right]\\
& \quad \times \exp \left[ - \sum_{q=1}^{K} \Lambda_{q}(T | Z_j) \right],
\end{aligned}
\end{eqnarray}
where $\mathbb{I}(\delta_{ji}=q)$ represents the indicator function aforementioned and $\boldsymbol\theta = \left( \boldsymbol\beta, \boldsymbol\alpha \right)$ with $\boldsymbol{\beta}=(\beta_1, \dots, \beta_K)$ and $\boldsymbol{\alpha}=(\alpha_1, \dots, \alpha_K)$; for $i=1,\dots,n_{j}$; $j=1, \dots, m$ and $q=1, \dots, K$. Thus, the overall likelihood function is represented by
\begin{equation}\label{verossim3}
\begin{aligned}
L( \boldsymbol\theta, \textbf{Z} \mid \boldsymbol{D} )&= \prod_{j=1}^{m} L_j( \boldsymbol\theta, Z_j | D_j )  \\
      &= c \prod_{j=1}^{m} Z_j^{n_j} \prod_{q=1}^{K} \left[ \beta_q^{n_{jq}} \alpha_q^{ n_{jq} \beta_q } T^{- n_{jq} \beta_q } \prod_{i=1}^{n_{jq}}  \left( t_{ji}^{\beta_q} \right)  \right] \\
      & \quad \times \exp\left[ - Z_j \sum_{q=1}^{K} \alpha_q \right] \\
      &\propto \prod_{j=1}^{m} Z_j^{n_j} \prod_{q=1}^{K} \gamma(\beta_{q}\mid n_{q} + 1, n_{q}\hat{\beta}_q^{-1} ) \\
      & \quad \times \gamma( \alpha_{q} \mid n_{q} + 1, m), \\ 
 \end{aligned}
\end{equation}
where $n_j = \sum_{q=1}^{K} n_{jq}$; $n_q = \sum_{j=1}^{m} n_{jq} $; $n_{jq}=\sum_{i=1}^{n_j} \mathbb{I}(\delta_{ji}=q)$; $\prod_{i=1}^{n_{jq}} (\cdot) = \prod_{i=1}^{n_{j}} (\cdot)^{\mathbb{I}(\delta_{ji}=q) } $; $ c = \prod_{j=1}^{m} \prod_{i=1}^{n_{jq}} t_{ji}^{-1} $. In addition,
\begin{equation}\label{mlebetaq}
\hat{\beta}_q = n_q / \sum_{j=1}^{m} \sum_{i=1}^{n_{jq}} \log (T / t_{ji})
\end{equation}
is the MLE for $\beta_q$.

As indicated previously, the overall likelihood function (\ref{verossim3}) may be factored as a product of three quantities, as follows:
\begin{equation}\label{verossim4}
    L(\boldsymbol\theta,  \textbf{Z} \mid \boldsymbol{D} )=L_1(\textbf{Z}\mid \textbf{D})L_2(\boldsymbol\beta\mid \textbf{D})L_3(\boldsymbol\alpha\mid \textbf{D}),
\end{equation}
 where $L_1(\textbf{Z}\mid \textbf{D}) = \prod_{j=1}^{m} Z_j^{n_j}$; $L_2(\boldsymbol\beta\mid \textbf{D}) = \prod_{q=1}^{k} \gamma(\beta_{q}\mid n_{q} + 1, n_{q}\hat{\beta_q}^{-1} ) $ and $L_3(\boldsymbol\alpha\mid \textbf{D}) = \prod_{q=1}^{k}  \gamma( \alpha_{q} \mid n_{q} + 1, m)$  and it will be used later in our posterior analysis.

\section{Bayesian analysis}  \label{secIV}

This section, in turn, is divided into two parts. In the first,  we present the choice of the prior distributions for  $\beta_q$ and $\alpha_q$ ($q=1,\ldots,k$) in the PLP model. In this case, we consider a similar approach according to the study of \cite{bar1992bayesian}. In the second, we discuss a Bayesian nonparametric approach to model the uncertainty about the distribution of shared frailty. As we will see in this section, we can carry out an individual posterior analysis of the quantities of interest due to the orthogonality among $\alpha_q$ and $\beta_q$ and the assumption that $Z_j$s are stochastically independent of the failure processes $\lambda_q$s.

\subsection{Prior specification for $\alpha$ and $\beta$ }\label{PriorAlfaBetasection}

Selecting an adequate prior distribution using formal rules has been widely discussed in the literature \cite{kass1996selection}. In the repairable systems context, \cite{bar1992bayesian} considered the following class of prior for the PLP model 
\begin{equation}\label{barlevprior}
     \pi (\alpha, \beta) \propto \alpha^{-1} \beta^{-\zeta} ,
 \end{equation}
where $\zeta>0$ is a known hyperparameter. Following these authors, we apply their main results in the setting of repairable systems under competing risks using the particular parametric formulation of PLP (\ref{plpCRZmulti}). Thus, we propose the prior distribution for the referred context as follows:
\begin{equation}\label{jointpriormodel1}
\pi(\boldsymbol{\alpha, \beta} ) \propto \prod_{q=1}^{K} \alpha_q^{-1} \beta_q^{-\zeta}.
\end{equation}

This class of prior distributions includes the invariant Jeffreys' prior when $\zeta=1$. Moreover, it reduces to (\ref{barlevprior}) when $q=1$. Further, we will discuss the chosen value for $\zeta$, and necessary conditions for the obtained posterior to be proper.

Note that, due to (\ref{verossim4}) and the assumption that $Z_j$s are stochastically independent of the failure processes $\lambda_q$s, the joint posterior distribution of (\ref{jointpriormodel1}) is proper. Note also that, the marginal distributions $\pi(\boldsymbol\beta \mid \textbf{D})$ and $\pi(\boldsymbol\alpha \mid \textbf{D})$ are proper since they are independent, as follows:
\begin{equation}\label{margpriorsalfbet}
\begin{aligned}
\pi(\boldsymbol\beta \mid \textbf{D}) = \prod_{q=1}^{k} \gamma\left(\beta_{q}\mid n_{q} +1-\zeta, n_{q}\hat{\beta}_q^{-1} \right)  
\end{aligned}
\end{equation}
and
\begin{equation}\label{margpriorsalfbet2}
\begin{aligned}
\pi(\boldsymbol\alpha \mid \textbf{D}) = \prod_{q=1}^{k}  \gamma( \alpha_{q} \mid n_{q} , m). 
\end{aligned}
\end{equation}

Since $\pi(\boldsymbol\alpha \mid \textbf{D})$ is the product of independent gamma distributions, then the marginal joint distribution $\pi(\boldsymbol\alpha \mid \textbf{D})$ is proper. Using the same idea, $\pi(\boldsymbol\beta \mid \textbf{D})$ is the  product of independent gamma distributions if $n_{q}>\zeta$ and, therefore, is a proper marginal posterior distribution.

This work adopts the quadratic loss function, hence the Bayes estimator is the posterior mean which has optimality under Kullback-Leibler divergence. It is worth pointing out that, in this chapter, the notation adopted for posterior mean will be $\hat{\alpha}_{q}^{Bayes}$ and $\hat{\beta}_{q}^{Bayes}$. Therefore,
 \begin{eqnarray}\label{Bayesest}
& & \hat{\alpha}_{q}^{Bayes} =  \mathbb{E}(\alpha_{q} \mid \textbf{D})  = \frac{n_q}{m}  \nonumber\\
& &\hat{\beta}_{q}^{Bayes} =  \mathbb{E}(\beta_{q} \mid \textbf{D}) = \frac{(n_{q} +1 - \zeta)}{n_{q}}\hat{\beta}_{q}. 
\end{eqnarray}

Besides the good properties mentioned above, we have that
\begin{equation}\label{expectd}
\begin{aligned}
&\mathbb{E}\left[\hat{\alpha}_{q}^{Bayes}\right]=\alpha_q \ \ \mbox{and} \\&
\mathbb{E}\left[\hat{\beta}_{q}^{Bayes}\right]=\mathbb{E}\left[\frac{(n_{q} +1- \zeta)}{n_{q}}\hat{\beta}_{q}\right]=\beta_q \quad \mbox{if} \quad \zeta=2.
\end{aligned}
\end{equation}

Therefore, assuming that $\zeta=2$ we have that both $\hat{\alpha}_{q}^{Bayes}$ and $\hat{\beta}_{q}^{Bayes}$ are unbiased estimators for $\alpha_q$ and $\beta_q$.

\subsection{Bayesian nonparametric approach for frailty distribution }\label{fragilitydistr}

This work presents the frailty distribution as an unknown distribution, therefore we will apply the Bayesian nonparametric methodology. Traditionally, the key idea of the Bayesian nonparametric approach is to obtain inference on an unknown distribution function using process priors on the spaces of densities. According to a definition provided by \cite{sethuraman1994constructive}, the nonparametric Bayesian model involves infinitely many parameters. To better understand the technical definition of Bayesian nonparametric models in a broad way, please see \cite{dey2012practical}, \cite{antoniak1974mixtures}, for example. There are many methods that specify more flexible density such as finite mixtures, DP, DPM, and mixture of Polya trees. Here, we considered DPM for logarithm of the frailty  $\textbf{W} = \log( \boldsymbol{Z} )$, represented by
\begin{eqnarray}\label{dpm1}
W_1,\dots, W_m & \sim  & F \nonumber \\
F & \sim & \mathcal{D}(c,F_{0}),
\end{eqnarray}
where $\mathcal{D}$ is the DP prior with base distribution  $F_0$; $c$ is the concentration parameter and $\textbf{W}=(W_1, \dots, W_m)'$. {\em c} can also be interpreted as a precision parameter that indicates how close the $F$ distribution  is to the base distribution $F_0$ \cite{escobar1995bayesian}.

Using the stick-breaking representation discussed in \cite{sethuraman1994constructive}, a DPM of Gaussian distribution can be represented as infinite mixtures of Gaussian, which is an extension of the finite  mixture model. Therefore, a density function of $W$ can be represented by
\begin{equation}\label{npbDPM1}
f_{W}(W) = f_{W}(W\mid \boldsymbol\Omega ) = \sum_{l=1}^{\infty} \rho_{l} \mathcal{N}(w \mid \mu_{l}, \tau_{l}^{-1} ),
\end{equation}
where $\mathcal{N}( \cdot \mid \mu, \tau^{-1} )$ denotes a normal density function with parameters $(\mu, \tau^{-1})$; $\boldsymbol\Omega = \{ \boldsymbol\rho,\boldsymbol\mu,\boldsymbol\tau\}$ is the infinite-dimensional parameter vector describing the mixture distribution for $W$;  $ \boldsymbol\rho = \{\rho_{l} \}_{l=1}^{\infty}$ represents the vector of weights, $\boldsymbol\mu = \{\mu_{l} \}_{l=1}^{\infty}$ is the vector of means and $\boldsymbol\tau = \{\tau_{l} \}_{l=1}^{\infty}$ is the vector of precision, for $l = 1,2,\dots$. Note that the density function of $Z$ can be calculated as follows:
\begin{eqnarray}\label{npbDPM2}
f_{Z}(Z) = f_{Z}(Z\mid \boldsymbol\Omega )& = & \sum_{l=1}^{\infty} \rho_{l} \mathcal{LN} (z \mid \mu_{l}, \tau_{l}^{-1} ),
\end{eqnarray}
where  $\mathcal{LN}(\cdot \mid \mu, \tau^{-1} )$ denotes log-normal density functions with parameters $\mu$ and  $\tau^{-1}$. Therefore, $Z$ can be represented as the infinite mixture log-normal. Note that the base distributions of $Z$ and $W$ are a log-normal and a normal distribution, respectively.  

\subsubsection*{Prior specification for $\Omega$ }\label{hmcsection}

As shown before, $\boldsymbol{\Omega}$ represents a collection of all unknown parameters in  (\ref{npbDPM1}) and (\ref{npbDPM2}). Based on this, we specified a prior distribution for $\boldsymbol{\Omega}$ as follows. Firstly, we specify a prior for $\boldsymbol{\rho}$.

Using the stick-breaking representation for prior distribution of $\boldsymbol\rho$, denoted by $\pi(\boldsymbol\rho)$, parameter vector $\boldsymbol\rho $ is reparameterized as follows:
\begin{eqnarray}\label{npbnu}
\rho_{1} & = & \nu_{1}, \nonumber \\
\rho_{l} & = & \prod_{o=1}^{l-1}(1 -\nu_{o})\nu_{l}, ~ \forall \ o =2,3,\ldots,
\end{eqnarray}
where the prior distribution of the vector $\boldsymbol\nu = \{\nu_{l} \}_{l=1}^{\infty} $ is independent and identically distributed with beta distribution denoted by \begin{equation}\label{priorNu}
    \boldsymbol{\nu} \sim \mathcal{B} (1,c),
\end{equation}
and the hyper-prior distribution of $c$ is 
\begin{equation}\label{priorCpar1}
    \pi(c) \sim \mathcal{G}(ac_0 , bc_0),
\end{equation}
where $\mathcal{G}(\cdot, \cdot)$ represents the gamma distribution \cite{escobar1995bayesian}. Besides, we chose a normal-gamma distribution as the prior of $(\mu_l,\tau_l) \sim \mathcal{NG}(m_0,s_0,d_0p_0,d_0)$,  for $l = 1,2,\ldots$, due to the fact that this prior is conjugate to the normal distribution, where
\begin{eqnarray*}
& & \mu_{l} \mid \tau_l \sim \mathcal{N}(m_0, (s_0 \tau_l)^{-1}) \, ,\\
& & \tau_{l} \sim \mathcal{G}(d_0, d_0p_0) \,.
\end{eqnarray*}

Thus, joint prior density of $\boldsymbol{\Omega}$ can be expressed as
\begin{equation}\label{priortOmega}
\pi(\boldsymbol{\boldsymbol\Omega}) = \pi(c)\pi(\boldsymbol{\rho}) \pi(\boldsymbol{\mu},\boldsymbol{\tau}).    
\end{equation}

For our Bayesian estimation scheme, the joint posterior distribution of $\textbf{Z}$ and all the unknown parameters in $\boldsymbol\Omega$ are reached by joining all the prior information (\ref{npbDPM2}), (\ref{priortOmega}) and the likelihood function (\ref{verossim4}), as follows:
\begin{eqnarray}\label{postZ}
\pi(\textbf{Z},\boldsymbol\Omega \mid \textbf{D})  \propto  L_1(\boldsymbol{Z}|\textbf{D}) f_{Z}(\textbf{Z} \mid \boldsymbol\Omega)  \pi(\boldsymbol{\boldsymbol\Omega}).
\end{eqnarray}

However, it is easy to see that (\ref{postZ}) does not have a closed form. Besides, the marginal posterior of $\textbf{Z}$ is intractable and it is therefore necessary to use MCMC algorithms, as we will see next. Recalling that one of our primary goals is to estimate $Z_j$s, thus, the Bayes estimator of $\textbf{Z}$ is given by 
\begin{eqnarray}\label{BayesestZ}
& & \hat{\textbf{Z}}^{Bayes} =  \sum_{i=1}^{L}\frac{\textbf{Z}^{(i)}}{L},
\end{eqnarray}
 where $\textbf{Z}^{(i)}$ is the $i$-th iteration  and $L$ is the total number of iterations of the MCMC chain.

\subsection*{MCMC algorithm}\label{hmcsection2}
This section describes an MCMC algorithm to sample from the posterior distribution of $Z$. Our algorithm is based on \cite{kalli2011slice}, and its main characteristic is to estimate infinite parameters by introducing latent variables. We introduce a finite set of latent variables with uniform distribution with parameters 0 and 1, denoted by   $U \sim \mbox{Uniform}[0,1]$. Therefore, applying the variable $U$ in (\ref{npbDPM2}) follows the joint density of $(Z, U)$
\begin{equation}\label{mc1}
  f_{Z,U}(z, u \mid \boldsymbol\Omega ) = \sum_{l=1}^{\infty}  \mathcal{LN}(z\mid \mu_{l}, \tau_{l}^{-1}) \mathbb{I}(u < \rho_{l}),
\end{equation}
where $\mathbb{I}(\cdot)$ is an indicator function. Note that there is a finite number of elements in $\boldsymbol\rho$ which are greater than $u$, denoted as $A_{\boldsymbol\rho}(u) = \{j : \rho_{j}> u\}$. Therefore, the representation in (\ref{mc1}) is similar to
\begin{equation}\label{mc2}
  f_{Z,U}(z, u \mid \boldsymbol\Omega ) = \sum_{l \in A_{\rho}}  \mathcal{LN}(z \mid \mu_{l}, \tau_{l}^{-1}),
\end{equation}
so that, given $\textbf{U}$, the number of mixture components is finite for $\textbf{Z}$.

In order to simplify the likelihood, we  introduce a new discrete latent variable $Y$ which indicates the mixture component that $Z$ comes from
 \begin{equation}\label{mc3}
  f_{Z,U,Y}(z, u, Y = l \mid \boldsymbol\Omega ) =  \mathcal{LN}(z\mid \mu_{l}, \tau_{l}^{-1}) \mathbb{I}(l \in  A_{\boldsymbol\rho}(u)).
\end{equation}
Note that $Pr(Y = l \mid \boldsymbol\Omega) = \rho_{l}$,  $\forall l = 1,2,\ldots $, therefore the conditional distribution of  $Z \mid U, Y = l$ is log-normal with parameters $\mu_{l}$  and $\tau_{l}^{-1}$, so $W\mid U, Y = l \sim \mathcal{N}(\mu_{l},\tau_{l}^{-1})$. Hence, the complete posterior distribution of $\textbf{Z}, \boldsymbol\Omega$ with the latent variables $\textbf{U}$ and $\textbf{Y}$ is given by
\begin{equation}\label{mc4}
\begin{aligned}
\pi(\textbf{Z},\boldsymbol\Omega,\textbf{U},\textbf{Y} \mid \textbf{D}) & \propto  L_1(\boldsymbol{Z}|\textbf{D}) f_{Z,U,Y}(\textbf{Z} \mid \boldsymbol\Omega,\boldsymbol{U},\boldsymbol{Y} ) f_U(\boldsymbol{U}) \\
& \quad \times Pr(\boldsymbol{Y}\mid \boldsymbol\Omega) \pi(\boldsymbol{\boldsymbol\Omega}),
\end{aligned}
\end{equation}
where $\boldsymbol{U} = \{U_{j}\}_{j=1}^{m} $ and  $\boldsymbol{Y} = \{Y_{j}\}_{j=1}^{m} $ are latent variables.

\subsubsection*{Hybrid MCMC - computational strategy}
Using the latent variables presented above, we now construct the following MCMC algorithm which is a combination of the Gibbs sampler with the HMC method. For more details on the HMC method, see \cite{neal2011mcmc}. We chose the HMC algorithm because it generates samples with less dependence with a high probability of acceptance between state if compared with the Random Walk Metropolis-Hastings algorithm. The Gibbs algorithm requires knowledge of complete conditional distributions in order to be able to sample from them. For further details, see \cite{kalli2011slice} and \cite{escobar1995bayesian}. The complete conditional distributions are listed below.


\begin{itemize}

\item[1.] \textbf{Conditional Distribution of c}

\cite{escobar1995bayesian} shows that given $\textbf{Y}$, the parameter is independent of all other parameters and the conditional distribution of $c$ is given by
\begin{equation}\label{conddistc1}
\begin{aligned}
\pi( c \mid \textbf{Y}) &\propto  (c + m)c^{y^*-1}\mathcal{G}\left(c\mid ac_0, bc_0 \right)\mathbb{B}(c+1,m)\\ 
& \quad \times \mathbb{I}(c > 0 ),
\end{aligned}
\end{equation}
where $y^* = \max(\textbf{Y})$ and $\mathbb{B}(\cdot,\cdot)$ is the Beta function. Using the definition  of the Beta function  we can create an auxiliary variable $\xi$ with the joint distribution for which the marginal distribution is (\ref{conddistc1}) and is given by
\begin{equation}\label{conddistc2}
\begin{aligned}
\pi( c,\xi \mid \textbf{Y}) &\propto  (c + m)c^{y^*-1}\xi \mathcal{G}\left(c\mid ac_0, bc_0 \right) \\
& \quad \times \xi^c\left(1-\xi\right)^{m-1} \mathbb{I}(c > 0 )\mathbb{I}(0 < \xi < 1 ).
\end{aligned}
\end{equation}
 Hence, it follows that the conditional posteriors of $\xi$ and $c$  are given by
\begin{equation}
\label{conddistc3}
\xi \mid c, \textbf{Y} \sim \mathcal{B}(c+1, m)
\end{equation}
and
\begin{equation}
\label{conddistc4}
 c \mid \xi,  \textbf{Y} \sim p_{\xi}\mathcal{G}(a_{1}^{*} ,b_{1}^{*}) +  (1 - p_{\xi}) \mathcal{G}(a_{2}^{*},b_{1}^{*}),
\end{equation}
where $a_{1}^{*} = a_{0} + y^*$,  $a_{2}^{*} = a_{1}^{*} + 1$, $b_{1}^{*}= b_0 - \log(\xi)$ and  $p_{\xi} = (a_{0} + y^* - 1)/(a_{0} + z^* - 1 + m(b_0 - \log(\xi)$. Therefore, $c$ can be sampled using the auxiliary $\xi$ with equations (\ref{conddistc3}) and (\ref{conddistc4}).

\item[2.] \textbf{Conditional Distribution of $\boldsymbol\nu$}

Note that by equations (\ref{mc3}) and (\ref{mc4}),  $\nu$ depends on $\textbf{Y}$, $\textbf{U}$ and $c$, therefore the conditional distribution of $\nu$ is
\begin{footnotesize}
\begin{equation}\label{conddistnu}
\begin{aligned}
\nu_{l} \mid \textbf{Y}, \textbf{U}, c & \sim \\
& \left\{  \begin{array}{ll}
   \mathcal{B}(n_{l}+1,m + \sum_{o=1}^{l} n_{o} + c) &, \forall l = 1,\ldots, y^{*} \\
   \mathcal{B}( 1, c) &, \forall l =  y^{*}+1, \\ 
   & \quad y^{*}+2,\ldots ,  \\
  \end{array} \right.
\end{aligned}
\end{equation}
\end{footnotesize}
where $n_{l}$ is the number of observations in the {\em l}-th component.
It is worth noting that in order to sample $\boldsymbol\rho$ it is enough to simulate $\boldsymbol\nu$ calculated by equation (\ref{npbnu}).

\item[3.] \textbf{Conditional Distribution of $\textbf{U}$}

The latent variable $U$ depends only on $\boldsymbol\rho$, and the conditional distribution of $ \textbf{U}$ is
\begin{equation}\label{conddistu}
U_{j} \mid \boldsymbol\rho \sim  \mbox{Uniform}[0, \rho_j] \ \forall j = 1,2,\ldots, m . 
\end{equation}

\item[4.] \textbf{Conditional Distribution of $\boldsymbol\mu$ and $\boldsymbol\tau$}

The  $\mu$ and $\tau$ parameters of each component are independent and adding the fact that the Normal-Gamma is conjugated from the Normal distribution, the conditional distribution of $\boldsymbol\mu$ and $\boldsymbol\tau$ is given by
\begin{footnotesize}
\begin{equation}\label{conddistmu}
\begin{aligned}
\mu_{l}, \tau_{l} \mid \textbf{Y} \sim \left\{  \begin{array}{ll}
   \mathcal{NG}(m_l,s_l,d_lp_l,d_l) &, \forall l = 1,\ldots, y^{*} \\
   \mathcal{NG}(m_0,s_0,d_0p_0,d_0) &, \forall l =  y^{*}+1, y^{*}+2,\ldots ,  \\
  \end{array} \right.
\end{aligned}
\end{equation}
\end{footnotesize}
 where

$\begin{array}{ll}
   m_{l} &= \displaystyle\frac{s_0m_0 + n_l\bar{w}}{s_0 + n_l} \, ,\\
   s_{l} &= s_0 + n_{l}  \, , \\
   d_lp_l &= d_0p_0 + \displaystyle\sum_{j:y_j = l} (w_{j} -\bar{w} )^2 + \displaystyle\frac{s_0n_{l}}{s_0 + n_l} (m_0 -\bar{w})^2 \, ,\\
  d_l   &= d_0 + n_l \, ,\\
  \bar{w} & = \displaystyle\sum_{j:y_j = l} \frac{w_j}{n_l}.\\
  \end{array}$

\item[5.] \textbf{Conditional Distribution of Y} 

The latent variable Y is discrete, therefore using equations (\ref{mc3}) and (\ref{mc4}) the conditional distribution of  $\textbf{Y}$ is
\begin{equation}\label{conddistY}
Pr(Y_j = l \mid  \boldsymbol\Omega, \textbf{W}, \textbf{U}, \textbf{D}   ) \propto \mathcal{N}(w\mid \mu_{l}, \tau_{l}^{-1}) \mathbb{I}(l \in  A_{\rho}).
\end{equation}

\item[6.] \textbf{Conditional Distribution of Z} 

The conditional distribution of  $\textbf{Z}$ is given by
\begin{equation}\label{conddistZ}
\pi( \textbf{Z}\mid  \boldsymbol\Omega,\boldsymbol{U}, \boldsymbol{Y}, \boldsymbol{D} ) \propto 
  \prod_{j=1}^{m}   \mathcal{LN}(z_{j}\mid \mu_{Y_j},\tau_{Y_j}^{-1} )L_1(\boldsymbol{Z}|\textbf{D}),
\end{equation}
with restriction $ \bar{\textbf{Z}} =  1$. Different from the previous parameters and latent variable, we simulate them using the HMC algorithm. However, the HMC algorithm requires that the support random variable is unrestricted. Therefore, we transform the variable $\textbf{Z}$ to a variable with  unrestricted support as explained below.

Let $\textbf{Z}^{\ast}$ be a random vector with $m - 1$ elements and unrestricted support. We define the following variables:
\begin{eqnarray}
\label{Zrest}
 B_j & = &  \mbox{logit}^{-1}( {Z}_{j}^{\ast} - \log( m - j) ), \nonumber \\
 A_j & = &  \left(1 - \sum_{j^{'}=1}^{j-1} A_{j^{'}}\right)B_J \ \  \forall j = 1,2,\ldots, m-1, \nonumber \\
 A_m & = &  1 - \sum_{j^{'}=1}^{m-1} A_{j^{'}},
\end{eqnarray} 
where  $logit^{-1}$ is an inverse function of $logit$. Note that  the functions of transformed variables are bijection, $B_j  \in (0,1)$ and sum($\textbf{A}$) = 1.  Naturally, we assume that $\textbf{Z} = m \textbf{A}$. Therefore, the determinant of the Jacobian matrix is given by,
$$  \mid J(\textbf{z}^{\ast}) \mid  = \prod_{j=1}^{m-1}\left(b_{j}(1 - b_{j})\left(1 - \sum_{j^{'}=1}^{j-1} a_{j^{'}}\right)\right) . $$
Therefore, the conditional distribution of  $\textbf{Z}^{\ast}$ is given by
\begin{equation}\label{conddistZ1}
\begin{aligned}
\pi( \textbf{Z}^{\ast}\mid  \boldsymbol\Omega,\boldsymbol{U},\boldsymbol{Y}, \boldsymbol{D} ) & \propto 
  \mid J(\textbf{z}^{\ast}) \mid    \mathcal{LN}(z_{j}\mid \mu_{Y_j},\tau_{Y_j}^{-1} ) \\
  & \quad \times L_1(\boldsymbol{Z} | \textbf{D} ).
\end{aligned}
\end{equation} 
\end{itemize}

Thus, we constructed a Hybrid MCMC algorithm that combines Gibbs sampling with HMC sampling to sample $\textbf{Z}$ and $\boldsymbol\Omega$; see algorithm below.   

\begin{enumerate}
    \item[\textbf{1:}] Initialize $c^{(0)}$, $\textbf{Z}^{\ast(0)}$ and $\textbf{Y}^{(0)}$.
    \item[\textbf{2:}] Calculate $\textbf{Z}^{(0)}$ of Equation (\ref{Zrest}) and $\textbf{W}^{(0)} =  \log(\textbf{Z}^{(0)})$. 
    \item[\textbf{3:}] Draw $\xi^{(i)}$ from $\pi(\xi \mid  c^{(i-1)} , \textbf{Y}^{(i -1)} )$ of Equation (\ref{conddistc3}).
    \item[\textbf{4:}] Draw $c^{(i)}$  from  $\pi(c \mid \xi^{(i)} , \textbf{Y}^{(i -1)} )$ of Equation (\ref{conddistc4}).
    \item[\textbf{5:}] Draw $\nu_{l}^{(i)}$  from  $\pi(\nu_{l} \mid \textbf{Y}^{(i-1)},  c^{(i)} )$ of Equation (\ref{conddistnu}), $\forall l = 1,2,\ldots,y^{*}$.
    \item[\textbf{6:}] Calculate $\rho_{l}^{(i)}$ of Equation (\ref{npbnu}) $\forall l = 1,2,\ldots,y^{*}$.
    \item[\textbf{7:}] Draw $U_{j}^{(i)}$ from $ \pi(U_{j} \mid \boldsymbol\rho^{(i)}$)   of Equation (\ref{conddistu}) $\forall j = 1,2,\ldots,m$.
    \item[\textbf{8:}] Find the smallest $l^{\ast}$ such that $\sum_{l=1}^{l^{\ast}}\rho_{l} > ( 1 - \min(\textbf{U}^{(i)}))$ and  draw $\nu_{l}^{(i)}$  from  $\pi(\nu_{l} \mid \textbf{Y}^{(i-1)},  c^{(i)} )$ , $\forall l = y^{*}+1,\ldots, l^{\ast}$.
    \item[\textbf{9:}] Draw  $ \mu_{l}^{(i)}$ and $\tau_{l}^{(i)}$ from $ \pi(
\mu_{l}, \tau_{l} \mid \textbf{Y}^{(i-1)})$ of Equation (\ref{conddistmu}) $ \forall l = 1,2,\ldots, l^{\ast}$.
    \item[\textbf{10:}] Draw $ Y_j^{(i)}$ from $Pr(Y_j \mid  \boldsymbol\mu^{(i)} ,\boldsymbol\tau^{(i)}, \textbf{W}^{(i-1)}, \textbf{U}^{(i)}, \textbf{D}   )$ of Equation (\ref{conddistY})
$\forall j = 1,2,\ldots, m$.
    \item[\textbf{11:}] Draw $\textbf{Z}^{\ast(i)}$  from 
$\pi( \textbf{Z}^{\ast}\mid  \boldsymbol\mu^{(i)} ,\boldsymbol\tau^{(i)},\boldsymbol{U}^{(i)},\boldsymbol{Y}^{(i)})$  of Equation (\ref{conddistZ1}).
    \item[\textbf{12:}] Calculate $\textbf{Z}^{(i)}$ of Equation (\ref{Zrest}) and $\textbf{W}^{(i)} =  \log(\textbf{Z}^{(i)})$.
    \item[\textbf{13:}] Set $i = i +1$ and go to Step \#3.
\end{enumerate}

In this scheme, the HMC  sampler is applied in Step \#11. The algorithm was developed in the C++ language using the RccpArmadilho library \cite{RcppArmadillo}. Its main advantages are processing speed and interaction with the R program \cite{rsoftware}. This code was used both in the generation of posterior sampling and in the simulation study presented in the following section.

\section{Simulation study}  \label{secV}

In this section, a simulation study is performed to evaluate the efficiency of the Bayesian estimators via the Monte Carlo method. To make our presentation easier, we consider two causes of failure with distinct parameters for each cause $\boldsymbol\theta=(\beta_1, \alpha_1, \beta_2, \alpha_2)$. The proposed simulation design is consistent with the following setup: (i) there are $m=(10,50,100)$ systems, each observed on the fixed time interval from $(0,20]$; (ii) the failure process for each component follows a power-law NHPPs with intensity (\ref{plpCRZmulti}); (iii) among the many possible parameter choices, we provide details for ($\beta_1=1.2,  \alpha_1=5, \beta_2=0.7, \alpha_2=13.33$) and ($\beta_1=0.75,  \alpha_1=9.46, \beta_2=1.25, \alpha_2=12.69$);  and (iv) we generate each random observation $z_j$, $j=1,\dots,m$, {\em iid} with mean one and variance $\eta$, according to a gamma distribution. In addition, we consider a set of values for variance of $\boldsymbol{Z}$, $\eta = (0.5, 1, 5)$, indicating low, middle and high dependence degrees, respectively. For each setup of parameters, we obtain the mean number of failures (5, 13.3), (9.5, 12.7), respectively. In the first simulated scenario, the mean number of failures of one of the components is predominant over the other component. In the last scenario, the mean number of failures of each component are almost equal to each other. It is worth noting that the obtained results are similar for other parameter combinations and can be extended to more causes, i.e. $p>2$. Using the fact that the causes are dependent due to frailty term $Z_j$ and also using the known results from the literature about NHPPs \cite{rigdon2000statistical}, in each Monte Carlo replication the failure times and indicators of the cause of failure were generated as shown in the following algorithm.

\begin{enumerate}
    \item[\textbf{1:}] Generate iid $ z_j \sim \gamma(\eta, 1/ \eta)$ for $j=1, 2, \dots, m$, with mean one and variance $\eta$.
    \item[\textbf{2:}] For each cause of failure, 
generate random numbers $n_{j1}$ and  $n_{j2}$, $j=1,\ldots,m$, both from a Poisson distribution with mean $z_j \alpha_q$, for $ q = 1, 2 $, respectively.
    \item[\textbf{3:}] For the $q$-th cause of failure from $j$-th system, 
let the failure times be $t_{j,1,q} , \ldots , t_{j, n_j, q}$, 
where $t_{j, i, q} = T \, U_{ j, i,q}^{ 1 / \beta_{jq} }$ and 
$U_{j,1,q} , \ldots , U_{j,n_j , q}$ are the order 
statistics of a size $n_j$ random sample from 
the standard uniform distribution.
    \item[\textbf{4:}] Finally, to obtain the data 
in the form $(t_i , \delta_i)$, 
let the $t_i$s be the set of ordered 
failure times and set $\delta_i$ equal to $j$ 
according to the corresponding cause of failure 
(i.e., set $\delta_i =1$ if $t_i = t_{h,1}$ for some $h$ 
or $\delta_i =j$ depending on the cause of failure).
\end{enumerate}

\begin{table*}[!t]
\centering
\caption{The Bias, MSE, CP(95\%) from the estimates considering different values for variance of {\em Z} and number of systems ({\em m}) with scenario $\boldsymbol\theta$=(1.2, 5, 0.7, 13.3).}
\begin{tabular}{c|c|c|r|r|r|r|r}
\hline 
  $\eta$          &  Parameter & {\em m} & \multicolumn{1}{c|}{$\alpha_1$}   & \multicolumn{1}{c|}{$\alpha_2$}  & \multicolumn{1}{c|}{$\beta_1$} & \multicolumn{1}{c|}{$\beta_2$} & \multicolumn{1}{c}{$\eta$}  \\ \hline
           &            &  10 & -0.0041 & 0.0083 & -0.0001 & 0.0007 & 0.0784  \\ 
           &  Bias      &  50 & 0.0011 & -0.0055 & -0.0001 & 0.0002 & 0.0276  \\ 
           &            & 100 & -0.0001 & 0.0074 & 0.0000 & -0.0001 & 0.0165  \\ \cline{2-8} 
           &            &  10 & 0.7035 & 1.1696 & 0.0729 & 0.0882 & 0.2709  \\ 
0.5        &  MSE       &  50 & 0.3182 & 0.5093 & 0.0321 & 0.0390 & 0.1359 \\ 
           &            & 100 & 0.2230 & 0.3650 & 0.0225 & 0.0276 & 0.0957  \\ \cline{2-8} 
           &            &  10 & 0.9427 & 0.9443 & 0.9449 & 0.9501 & 0.8395  \\ 
           &  CP(95\%)  &  50 & 0.9483 & 0.9500 & 0.9459 & 0.9506 & 0.9366  \\ 
           &            & 100 & 0.9502 & 0.9496 & 0.9512 & 0.9488 & 0.9444  \\ \hline 
                     &            &  10 & 0.0084 & -0.0105 &  0.0003 & -0.0023 &  0.0307  \\ 
           &  Bias      &  50 & 0.0020 & -0.0007 & -0.0006 &  0.0001 &  0.0253  \\ 
           &            & 100 & 0.0010 & 0.0039 & 0.0000 & 0.0000 & 0.0158  \\ \cline{2-8} 
           &            &  10 & 0.6996 & 1.1449 & 0.0735 & 0.0879 & 0.5395  \\ 
1          &  MSE       &  50 & 0.3120 & 0.5185 & 0.0316 & 0.0393 & 0.2891  \\ 
           &            & 100 & 0.2231 & 0.3690 & 0.0226 & 0.0275 & 0.2015  \\ \cline{2-8} 
           &            &  10 & 0.9444 & 0.9517 & 0.9432 & 0.9477 & 0.9423  \\ 
           &  CP(95\%)  &  50 & 0.9532 & 0.9488 & 0.9477 & 0.9477 & 0.9544  \\ 
           &            & 100 & 0.9477 & 0.9472 & 0.9489 & 0.9492 & 0.9478  \\ \hline 
                     &            &  10 & 0.0174 & -0.0120 & -0.0005 &  0.0009 & -0.1693  \\ 
           &  Bias      &  50 & -0.0017 & -0.0085 & -0.0009 &  0.0000 &  0.0453  \\ 
           &            & 100 & 0.0000  & 0.0005 &  0.0001 & -0.0004 & -0.0425  \\ \cline{2-8} 
           &            &  10 & 0.7156 & 1.1508 & 0.0723 & 0.0873 & 2.2234  \\ 
5          &  MSE       &  50 & 0.3179 & 0.5141 & 0.0317 & 0.0390 & 2.1358  \\ 
           &            & 100 & 0.2239 & 0.3711 & 0.0223 & 0.0276 & 1.5038  \\ \cline{2-8} 
           &            &  10 & 0.9419 & 0.9508 & 0.9460 & 0.9483 & 0.9340  \\ 
           &  CP(95\%)  &  50 & 0.9476 & 0.9504 & 0.9500 & 0.9472 & 0.9473  \\ 
           &            & 100 & 0.9470 & 0.9462 & 0.9505 & 0.9476 & 0.9426  \\ \hline \hline 
\end{tabular}
\label{tableSimu1}
\end{table*}

Software R was used to implement this simulation study \cite{rsoftware}. We considered two criteria to evaluate the estimators' behaviour: 
the Bias, given by
$
\f{Bias}_{\hat{\theta}_i}=\sum_{j=1}^{M}(\hat\theta_{i,j}-\theta_i)/M 
$
and the MSE, given by
$
\f{MSE}_{\hat{\theta}_i} =  
\sum_{j=1}^{M}{(\hat\theta_{i,j}-\theta_i)^2}/{M},
$
where $M$ 
is the number of estimates 
(i.e.\ the Monte Carlo size), where we take 
$M=50,000$ throughout the section,
and $\boldsymbol{\theta}=(\theta_1,\ldots,\theta_p)$ is the vector of parameters. Additionally, we computed the $CP_{95\%}$. Good estimators should have Bias, MSE close to zero and adequate intervals should be short while showing  $CP_{95\%}$ close to 0.95. The Bias and MSE are widely used to measure the performance evaluation.

The Bayes estimators for $\beta_j$ and $\alpha_j$ were obtained using independent marginal posteriors according to gamma distributions given in (\ref{margpriorsalfbet}). Since the marginal posterior distributions for the parameters $\beta_j$ and $\alpha_j$ follow gamma distributions, we can obtain closed-form expressions for the posterior means and obtain the credibility intervals based on the 2.5\% and 97.5\% percentile posteriors. Hence, no MCMC was needed to obtain the estimates for these parameters. On the other hand, to obtain the estimates of the $Z_j$s, $j=1,\ldots,m$, we considered the HMC described in Section \ref{hmcsection2}. For each simulated data set, $10,000$ iterations were performed using the MCMC methods. As a burn-in, the first $5,000$ initial values were discarded. The Geweke criterion \cite{geweke1992evaluating} was considered to check the convergence of the obtained chains under a $95\%$ confidence level. In addition, trace and autocorrelation plots of the generated sampled values of each $Z_j$ showed that they converged to the target distribution. The remaining $5,000$ were used for posterior inference. Specifically, these values were used to compute the posterior means of $Z_j$s. Table \ref{tableSimu1} presents the Bias, the MSE and coverage probability with a $95\%$ confidence level of the Bayes estimates for $\alpha_1, \alpha_2, \beta_1, \beta_2$ and the variance of $Z$.

\begin{table*}[!t]
\centering
\caption{The Bias, MSE, CP(95\%) from the estimates considering different values for variance of {\em Z} and number of systems ({\em m}) with scenario $\boldsymbol\theta$=(0.75, 9.5, 1.25, 12.7)}
\begin{tabular}{c|c|c|r|r|r|r|r}
\hline 
  $\eta$          &  Parameter & m & \multicolumn{1}{c|}{$\alpha_1$}   & \multicolumn{1}{c|}{$\alpha_2$}  & \multicolumn{1}{c|}{$\beta_1$} & \multicolumn{1}{c|}{$\beta_2$} & \multicolumn{1}{c}{$\eta$}  \\ \hline
           &            &  10 & 0.0215 & -0.0189 & 0.0004 & 0.0000 & 0.0745  \\ 
           &  Bias      &  50 & 0.0017 & -0.0003 & 0.0004 & -0.0004 & 0.0218  \\ 
           &            & 100 & 0.0025 & -0.0005 & 0.0003 & 0.0001 & 0.0155 \\ \cline{2-8} 
           &            &  10 & 0.9632 & 1.1248 & 0.0787 & 0.1120 & 0.2691  \\ 
0.5        &  MSE       &  50 &  0.4341 & 0.4992 & 0.0347 & 0.0495 & 0.1312  \\ 
           &            & 100 & 0.3085 & 0.3569 & 0.0243 & 0.0350 & 0.0946  \\ \cline{2-8} 
           &            &  10 & 0.9498 & 0.9477 & 0.9467 & 0.9470 & 0.8346  \\ 
           &  CP(95\%)  &  50 & 0.9506 & 0.9516 & 0.9497 & 0.9508 & 0.9377  \\ 
           &            & 100 & 0.9471 & 0.9462 & 0.9509 & 0.9482 & 0.9417  \\ \hline 
                     &            &  10 & -0.0025 & -0.0013  & 0.0003  & 0.0003  & 0.0233  \\  
           &  Bias      &  50 & 0.0005 &  0.0039 & -0.0004 & 0.0003  & 0.0155  \\ 
           &            & 100 & 0.0024 & -0.0013  & 0.0001 & -0.0003  & 0.0087  \\ \cline{2-8} 
           &            &  10 & 0.9678 & 1.1311 & 0.0780 & 0.1138 & 0.5279  \\ 
1          &  MSE       &  50 & 0.4340 & 0.5065 & 0.0346 & 0.0497 & 0.2808  \\
           &            & 100 & 0.3087 & 0.3592 & 0.0246 & 0.0355 & 0.1987  \\ \cline{2-8} 
           &            &  10 & 0.9497 & 0.9503 & 0.9478 & 0.9471 & 0.9407  \\
           &  CP(95\%)  &  50 & 0.9495 & 0.9465 & 0.9477 & 0.9515 & 0.9510  \\ 
           &            & 100 & 0.9456 & 0.9470 & 0.9463 & 0.9482 & 0.9495  \\ \hline 
                     &            &  10 & -0.0148 & -0.0076 & -0.0008 & -0.0002 & -0.1558  \\
           &  Bias      &  50 & -0.0027 & -0.0039 &  0.0007 & -0.0001 &  0.0577  \\
           &            & 100 & 0.0040 & -0.0040 & -0.0003 &  0.0002 & -0.1141 \\ \cline{2-8} 
           &            &  10 & 0.9663 & 1.1197 & 0.0785 & 0.1119 & 2.2178  \\
5          &  MSE       &  50 & 0.4366 & 0.5044 & 0.0348 & 0.0497 & 2.0724  \\ 
           &            & 100 & 0.3075 & 0.3592 & 0.0246 & 0.0347 & 1.4197  \\ \cline{2-8} 
           &            &  10 & 0.9517 & 0.9512 & 0.9493 & 0.9491 & 0.9354  \\ 
           &  CP(95\%)  &  50 & 0.9491 & 0.9471 & 0.9499 & 0.9522 & 0.9479  \\
           &            & 100 & 0.9489 & 0.9467 & 0.9497 & 0.9551 & 0.9485  \\ \hline \hline 
\end{tabular}
\label{tableSimu2}
\end{table*}

As shown in Tables \ref{tableSimu1} and \ref{tableSimu2}, the biases of the Bayes estimator are very close to zero for all the parameters, while both Bias and MSE tend to zero as $m$ increases. Hence, in terms of Bias and MSE, the Bayes estimators provided accurate inferences for the parameters of the PLP model. In terms of coverage probabilities, we observed that using our Bayes estimators returned accurate credibility intervals even for a small number of system $m$. This result may be explained by the fact that our proposed Bayes estimators do not depend on asymptotic results to obtain the credibility intervals, which leads to accurate results for small sample sizes.

\section{Application to the warranty repair data}  \label{secVI}

\begin{figure}[!htb]
\centering
\includegraphics[scale=0.3]{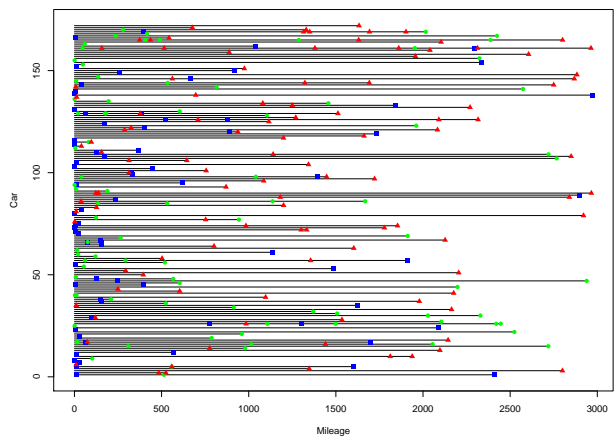}
\caption{Recurrences of three causes of failure for 172 cars from warranty claims data. The green circle represents the cause of failure 1, the red triangle represents the cause of failure 2 and the blue square represents the cause of failure 3.}\label{dotplot}
\end{figure}

The dataset considered in this section comprises the recurrent failure history of a fleet of identical automobiles obtained from a warranty claim database presented in \cite{somboonsavatdee2015parametric}. For the sake of clarity, our graphics present only the cars that presented failures in the observation period. Figure \ref{dotplot} shows the recurrence of failures of the 172 cars according to the cause of failure and the car mileage at each failure. The x-axis indicates the mileage. It is worth noting that the process of data collection has truncated time, where the observation period is 3000 miles for all cars. Each car from the fleet is represented by a horizontal line, where the cause of failure 1 is identified by the green circle, the cause of failure 2 by the red triangle and the cause of failure 3 by the blue square. 
We suppose that maintenance policy is minimal repair.

The main authors make only a table available (omitted here) containing the mileage to repeated failures of 172 vehicles, as well as the associated cause of failure. There were 76 failures related to the cause of failure 1, 87 related to the cause of failure 2 and 111 related to the cause of failure 3. They also pointed out that there were 267 cars that did not fail during the observation period. However, following the correct methodology, we consider 439 automobiles in our analysis.

\begin{figure}[!htb]
\centering
\includegraphics[scale=0.45]{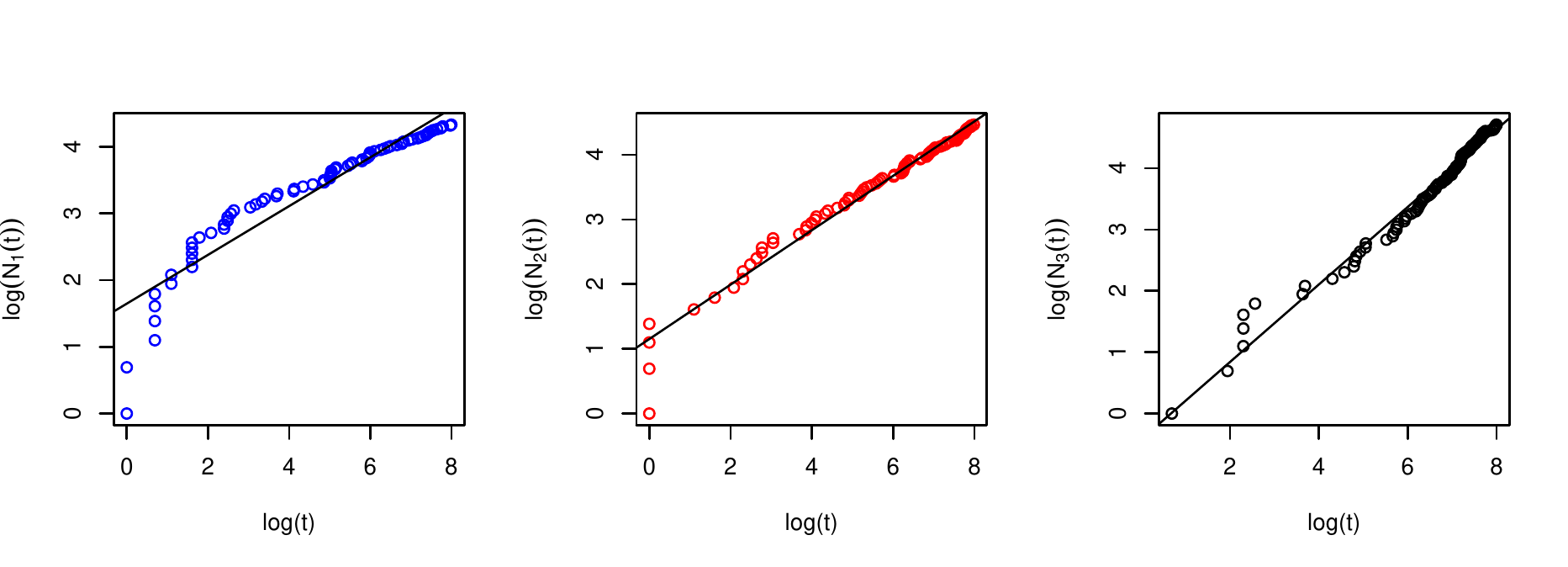}
\caption{The plot shows a fairly linear pattern for the three causes of failure indicating the fit according to the PLP model: cause 1 (blue circles); cause 2 (red circles) and cause 3 (black circles)}\label{duane1}
\end{figure}

Following \cite{somboonsavatdee2015parametric, somboonsavatdee2015statistical}, we assessed the adequacy of the PLP for each cause of failure using the Duane plot \cite{duane1964learning, crow1974reliability,rigdon2000statistical}. Figure \ref{duane1} shows plots of logarithm of the number of failures $N_q(t)$ (for $q=1,2,3$) against the logarithm of the accumulated mileage at failure. Since the three plots exhibit reasonable linearity, the PLP model seems to be adequate.

Since the PLP is adequate we consider our proposed approach to fit the data. As presented in Section \ref{PriorPostsection}, we assume the prior distribution (\ref{jointpriormodel1}) for parameters $\alpha_q$ and $\beta_q$ ($q=1, 2, 3$) and, consequently, the marginal posterior distributions (\ref{margpriorsalfbet}). On the basis of the latter consideration, the posterior mean estimates are computed in closed-form and the CIs are obtained directly from the gamma distribution. The results of the analysis are presented in Table \ref{table20}, which show Bayes estimates along with the corresponding SDs and CIs. According to these data, the estimates of the shape parameters $\left( \hat{\beta}_1, \hat{\beta}_2, \hat{\beta}_3 \right)$ are smaller than 1; see Table \ref{table20}. This clearly indicates improvement in reliability.
\begin{table}[!h]
\renewcommand{\arraystretch}{1.3}
\caption{Parameter estimates for warranty claim dataset}
\centering
\begin{tabular}{crcc}
\hline\hline
Parameter & Bayes & SD & CI (95\%) \\ 
\hline\hline
$\beta_1$ & 0.300 & 0.035 & [0.236 ; 0.372] \\ 
$\beta_2$ & 0.409 & 0.044 & [0.327 ; 0.500] \\ 
$\beta_3$ & 0.698 & 0.067 & [0.574 ; 0.835] \\ 
$\alpha_1$ & 0.173 & 0.020 & [0.136 ; 0.214] \\ 
$\alpha_2$ & 0.198 & 0.021 & [0.159 ; 0.242] \\ 
$\alpha_3$ & 0.253 & 0.024 & [0.208 ; 0.302] \\ 
$Var(Z)$ & 1.755 & 0.438 & [1.050 ; 2.777] \\ 
\hline \hline
\end{tabular}\label{table20}
\end{table}

The hybrid MCMC sampler algorithm presented in Section \ref{hmcsection2} was used to obtain a sample from the joint posterior distribution related to the frailty distribution. The initial values to start the sample of the chains for the DPM were random. For the MCMC chain, we considered 10,000 iterations initially, where the first 5,000 were discarded as burn-in samples and the last 5,000 iterations were used to compute the posterior estimates of $Var(Z)$ (at the bottom of the Table \ref{table20}) and the individual values of $Z_j$s, as presented in Figure \ref{conv2}. The convergence was monitored for the Geweke test assuming a $95\%$ confidence level (see Figure \ref{diagnostic1} in Appendix A). For completeness, we also present MCMC diagnostic plots, such as traces and autocorrelations for the HMC algorithm; see Appendix A.

It is worth pointing out that higher values of $Var(Z)$ signify greater heterogeneity among systems and more dependence between the times of the causes of failure for the same system. Therefore, as Table \ref{table20} shows, the posterior mean of $Var(Z)$ provides evidence of a meaningful dependence between the times of the causes of failure within a system.

\subsection{Insights on the unobserved heterogeneity}\label{insig}

As shown in Table \ref{table20}, the estimate of $Var(Z)$ shows that there is strong posterior evidence of a meaningful degree of heterogeneity in the population of systems. Table \ref{table30} (Appendix A) shows the estimated posterior means and the corresponding standard deviations of the $\hat{z}_j$s.

\begin{figure}[!htb]
\centering
\includegraphics[scale=0.45]{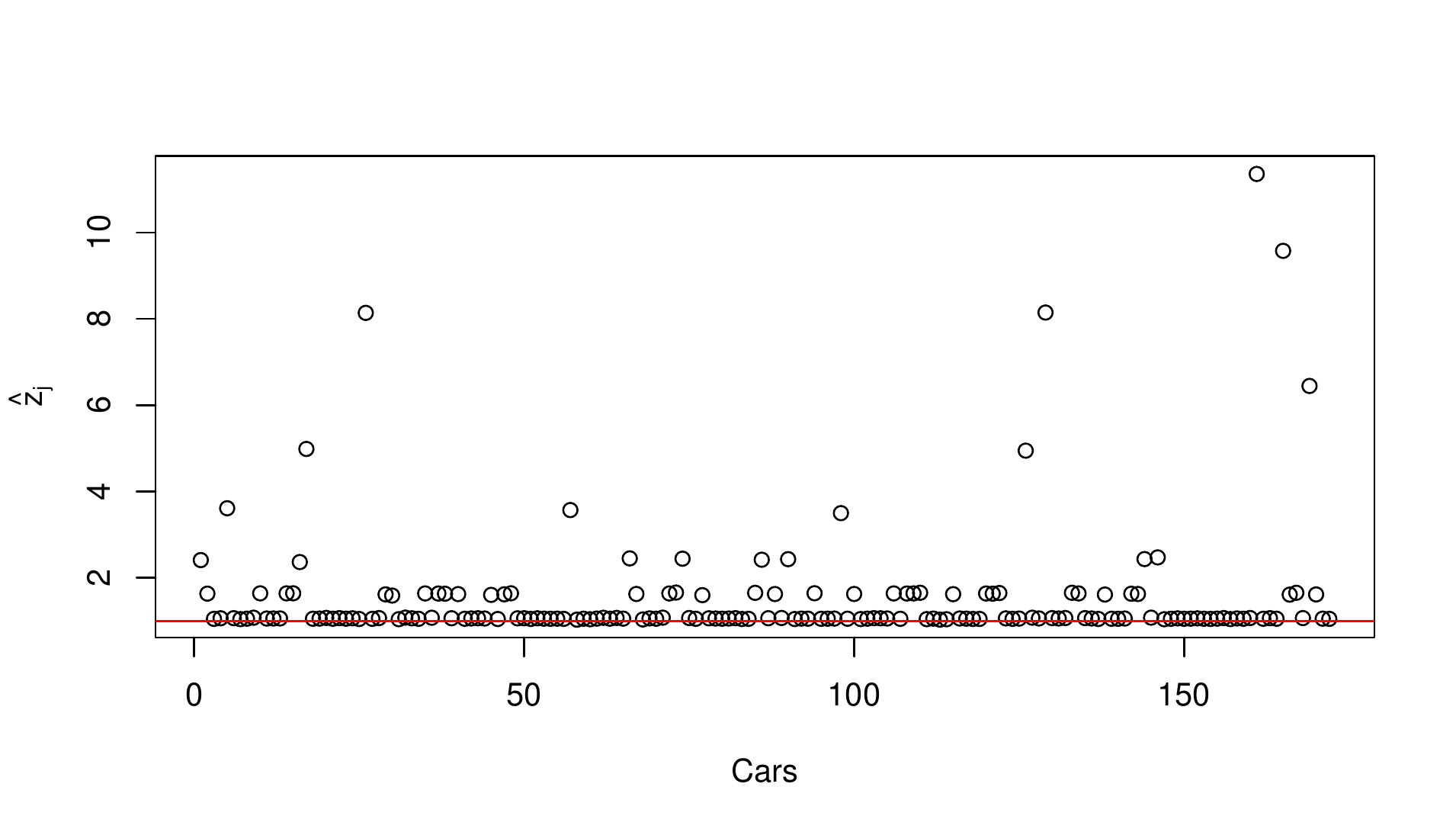}
\caption{The individual frailty estimates, $\hat{z}_j$'s. The red line highlights value 1 in the y-axis.}\label{conv2}
\end{figure}

Figure \ref{conv2} shows the individual frailty estimates (posterior means) of $\hat{z}_j, j=1,\ldots,172$. As mentioned earlier, each $Z_j$ acts in a multiplicative way in the specific-cause intensities. Thus it follows that values of $Z_j$ equal to or very close to 1 (red line) do not significantly affect such intensities. On the other hand, values larger than 1 indicate increased intensity. It is apparent that some cars have values of $Z_j$ greater than 2. These cars are probably subject to environmental stress variations or other unobserved issues, which make them more vulnerable than those with $Z_j$ values closer to or less than 1.

\begin{figure}[!htb]
\centering
\includegraphics[scale=0.45]{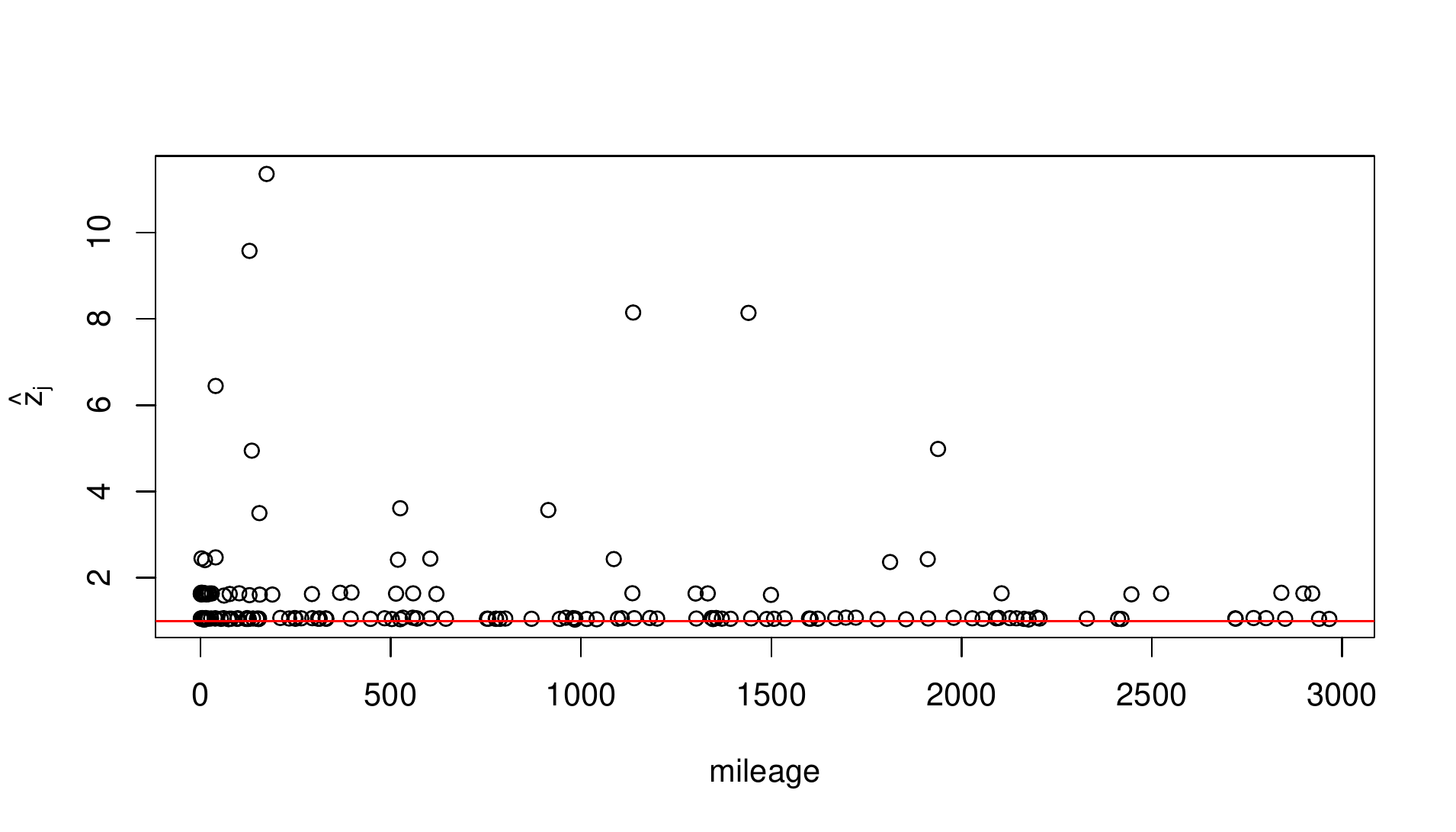}
\caption{Estimated frailty versus mileage observed at failure for each car in automobile warranty data. The red line highlights value 1 in the y-axis. The reasoning is that cars that are more frail failed earlier than ones that are less frail.}\label{carmilg}
\end{figure}

\begin{figure}[!htb]
\centering
\includegraphics[scale=0.45]{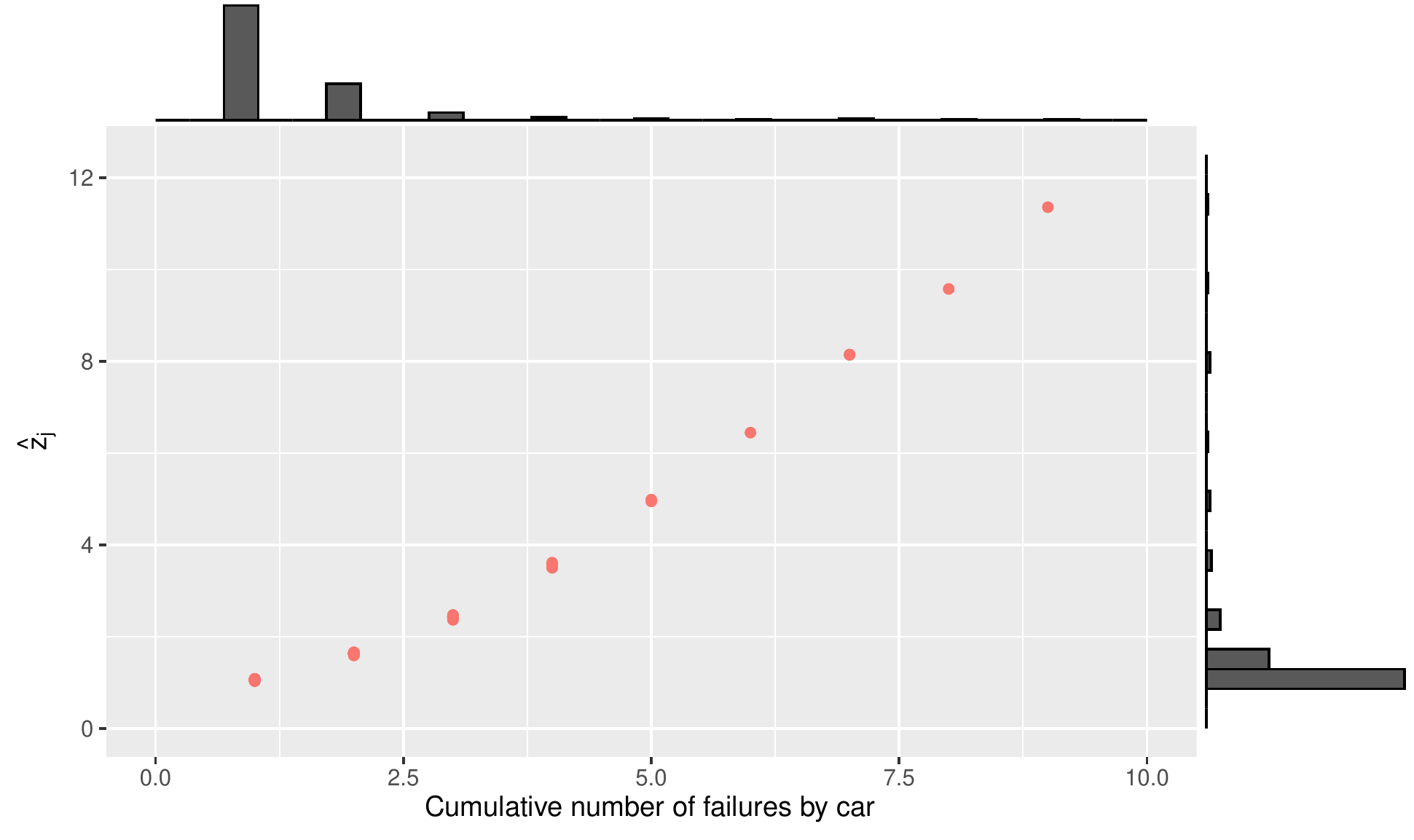}
\caption{Scatterplot of individual estimates $\hat{z}_j$ against cumulative number of failures by car. Note that systems with a large value of $\hat{z}_j$ experienced more failures than a system with a smaller value of $\hat{z}_j$.}\label{correlFN}
\end{figure}

Figure \ref{carmilg} indicates that the estimated frailties are overall larger for cars that had a failure early than those who had a failure later. We also note that a system with a large value of $\hat{z}_j$ experienced more failures than a system with a smaller value of $\hat{z}_j$ (see Figure \ref{correlFN}).

These outcomes indicate that neglecting these effects can result in an underestimation of the parameters. Overall, the multiplicative shared frailty model is appropriate for modeling this effect accurately.

\section{Conclusions}  \label{secVII}

In this work, we proposed a new approach to analyzing multiple repairable systems data under the action of dependent competing risks. We have shown how to model the frailty-induced dependence nonparametrically using a DPM which does not make restrictive assumptions about the density of the frailty variable.  Although some research has been carried out on nonparametric frailty in the reliability field \cite{slimacek2016nonhomogeneous, slimacek2017nonhomogeneous}, to the best of our knowledge, the proposed approach is the first for this competing risks setup. The main focus of this chapter was to provide estimates for the PLP model taking into account the dependence effect among component failures of the system. Such a dependence effect influences the statistical inferences of the model parameters, thus the misspecification of the frailty distribution may lead to errors when estimating the quantities of interest.

An orthogonal parametrization for the cause-specific intensity PLP parameters was presented, which allowed us to consider a generalized version of \cite{bar1992bayesian} prior distribution for the parameters of the model. Assuming the quadratic loss function as the risk function, we obtained the posterior mean for the parameters in closed-form expression. Moreover, since the marginal posterior distributions for the  PLP parameters follow gamma distributions, we obtained the credibility intervals directly for the quantile function. Assuming a specific value for $\zeta$, we obtained unbiased estimators for the cited parameters.
A simulation study was conducted to confirm our theoretical results, as well as to measure if the variability of the frailty distributions was correctly computed. This study returned excellent results that confirmed that our Bayes estimators are robust in terms of Bias, MSE and coverage probabilities.

Using nonparametric Bayesian methods with a mixture prior distribution enabled us to increase the amount of information beyond the parameter estimates. We considered a Bayesian nonparametric prior to describing the frailty distribution due to its flexibility in modeling unknown distributions. Although this model has infinite parameters, it is a flexible mixture model, parsimonious and straightforward to sample from. In this case, we chose the stick-breaking representation of the DP prior because of a simple implementation to build the algorithm.  Hence, we proposed a hybrid MCMC algorithm that comprises a mixture of the Gibbs sampler and the HMC method, thus generating a chain with little dependence.

The results of this investigation show that we can obtain more precise parameter estimations by considering the high flexibility due to nonparametric Bayesian prior density for $Z$. It also enables us to obtain insights into the heterogeneity between the systems by individually estimating $Z_j$s, as presented in Section \ref{insig}. The methodology proposed in this study may be of assistance to industrial applications and also where the interest may be in the phases of developmental programs of prototypes with purposes to predict the reliability, for example.

Our findings can be applied in real data sets based on the following assumptions. The proposed model requires $m$ identical repairable systems subjected to $K$ competing risks (assuming dependece). Minimal repair policy is assumed. The recurrent data structure (failure history) should be based on cause-specific intensity functions with PLP. The data sampling scheme (system observational period) is the time truncated case. Consider the shared frailty model to incorporate the dependence among the cause-specific recurrent processes. Finally, the dataset should be structured as Table \ref{tablemodel2} in Appendix C.

More flexible modeling can be further proposed by extending our approach to model the intensity function of failures of the NHPP nonparametrically since that the PLP intensity cannot capture non-monotonic behaviors. This extension would make the model more robust and flexible. In this case, we would have a fully nonparametric approach. The proposed study can also be further adapted under other types of repair such as perfect or imperfect. Our approach should be investigated further in these contexts.

\section*{Appendix A}\label{chapter:AppendixA}

In this appendix, we presented estimates of some $Z_j$'s associated to cars 1, 17, 26, 161, 165 and 169, according to Figure \ref{conv2} (these are the estimates that presented the highest values). For completeness, we also present here the Geweke diagnostic test for checking the convergence of the chains, as well as MCMC diagnostic plots, such as trace and autocorrelations for the HMC algorithm of some $Z_j$s.\newline

\begin{table}[!h]
\renewcommand{\arraystretch}{1.3}
\caption{Bayesian estimates of some $Z_j$s with their SD.}
\centering
\begin{tabular}{crc}
\hline
$Z_j$ & Bayes & SD \\ 
\hline
$Z_1$ & 2.469 & 1.711 \\ 
$\vdots$ & $\vdots$ & $\vdots$ \\ 
$Z_{17}$ & 5.1 & 2.87 \\ 
$\vdots$ & $\vdots$ & $\vdots$ \\ 
$Z_{26}$ & 8.269 & 3.669 \\ 
$\vdots$ & $\vdots$ & $\vdots$ \\ 
$Z_{161}$ & 11.519 & 4.359 \\ 
$\vdots$ & $\vdots$ & $\vdots$ \\ 
$Z_{165}$ & 9.941 & 4.038 \\ 
$\vdots$ & $\vdots$ & $\vdots$ \\ 
$Z_{169}$ & 6.615 & 3.354 \\
\hline
\end{tabular}\label{table30}
\end{table}

\newpage


\begin{figure}[!htb]
\centering
\includegraphics[scale=0.4]{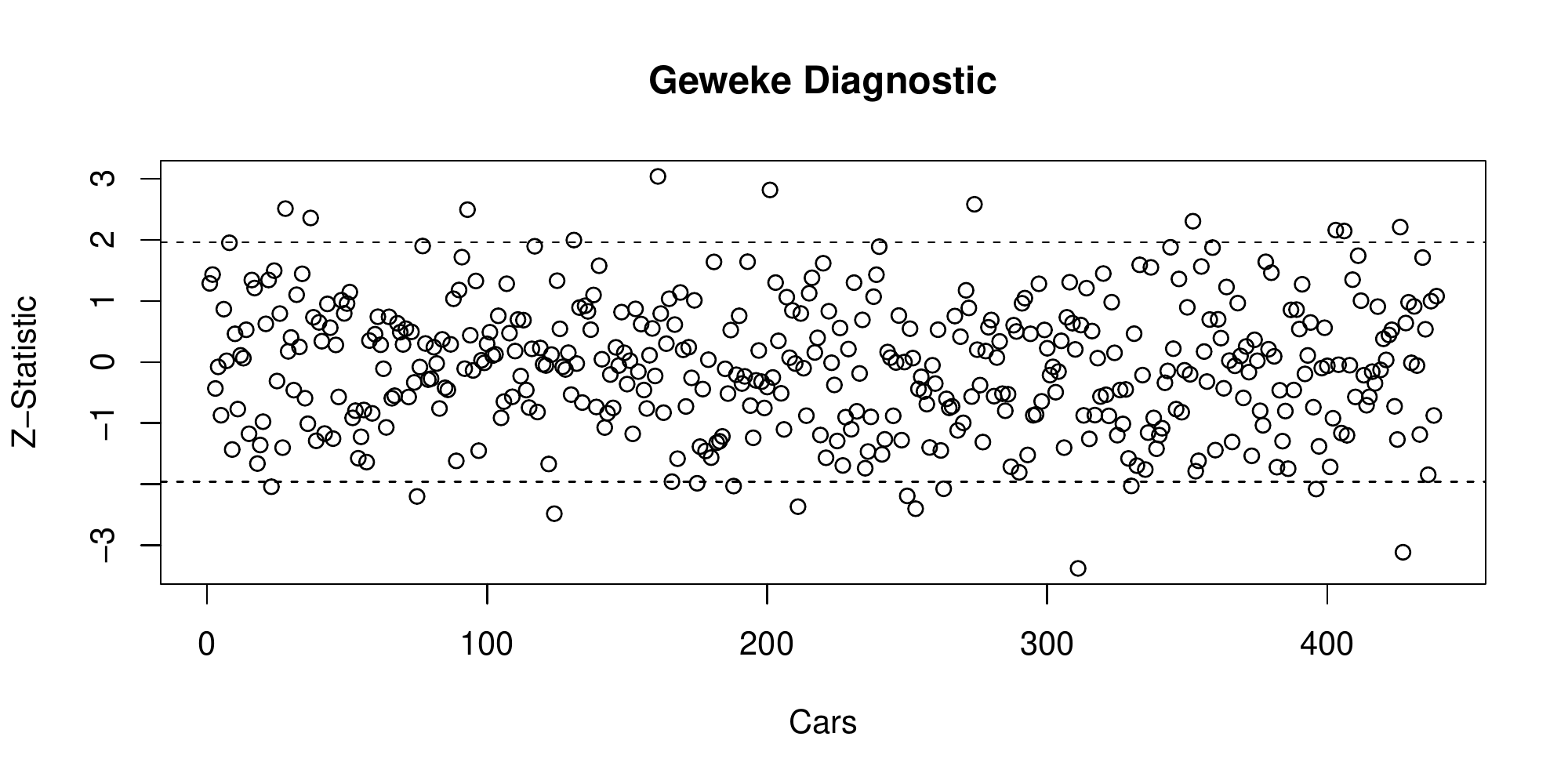}
\caption{Geweke diagnostic test - implemented using CODA package in R software.}\label{diagnostic1}
\end{figure}

\begin{figure}[!htb]
\centering
\includegraphics[scale=0.4]{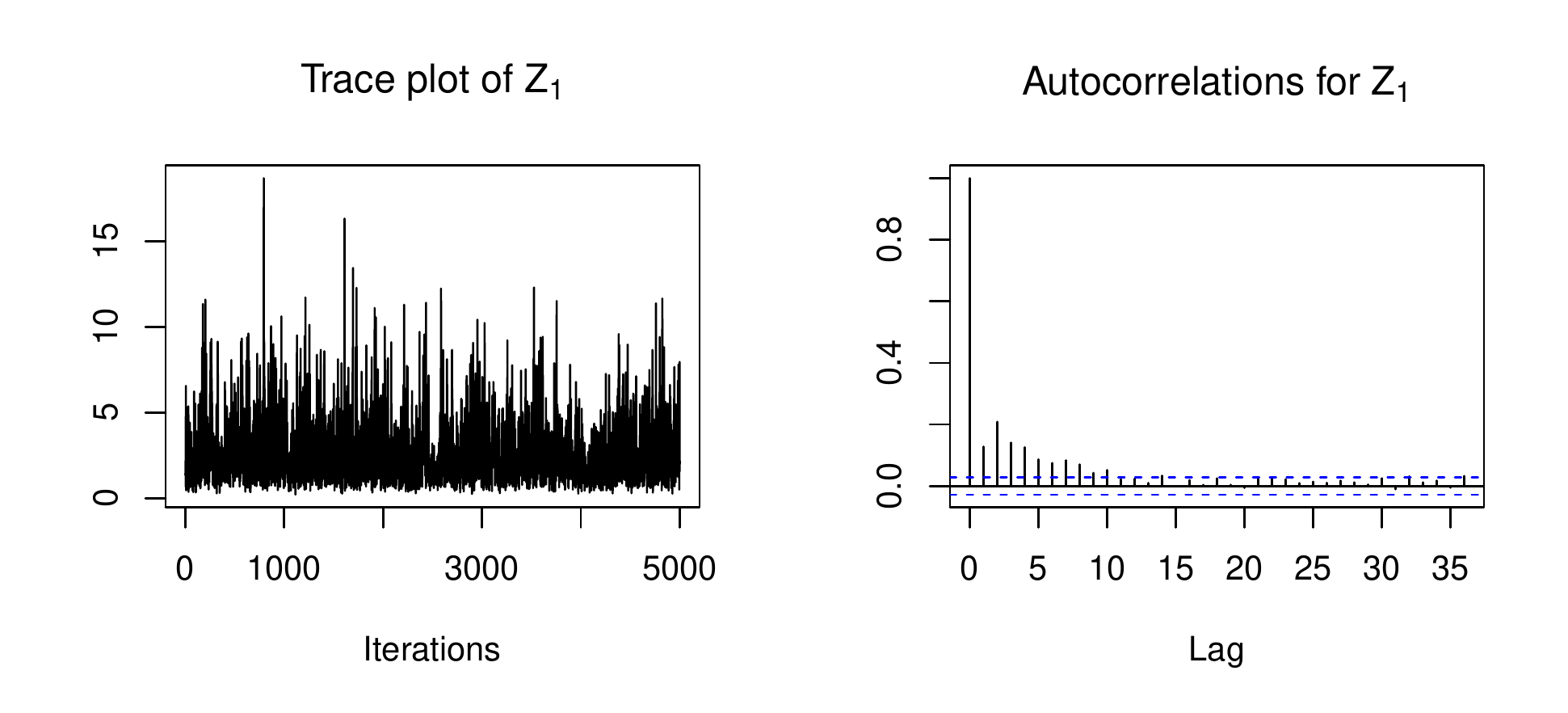}
\caption{Markov chain and autocorrelation plots for the HMC algorithm - $Z_1$.}\label{convZ1}
\end{figure}	

\begin{figure}[!htb]
\centering
\includegraphics[scale=0.4]{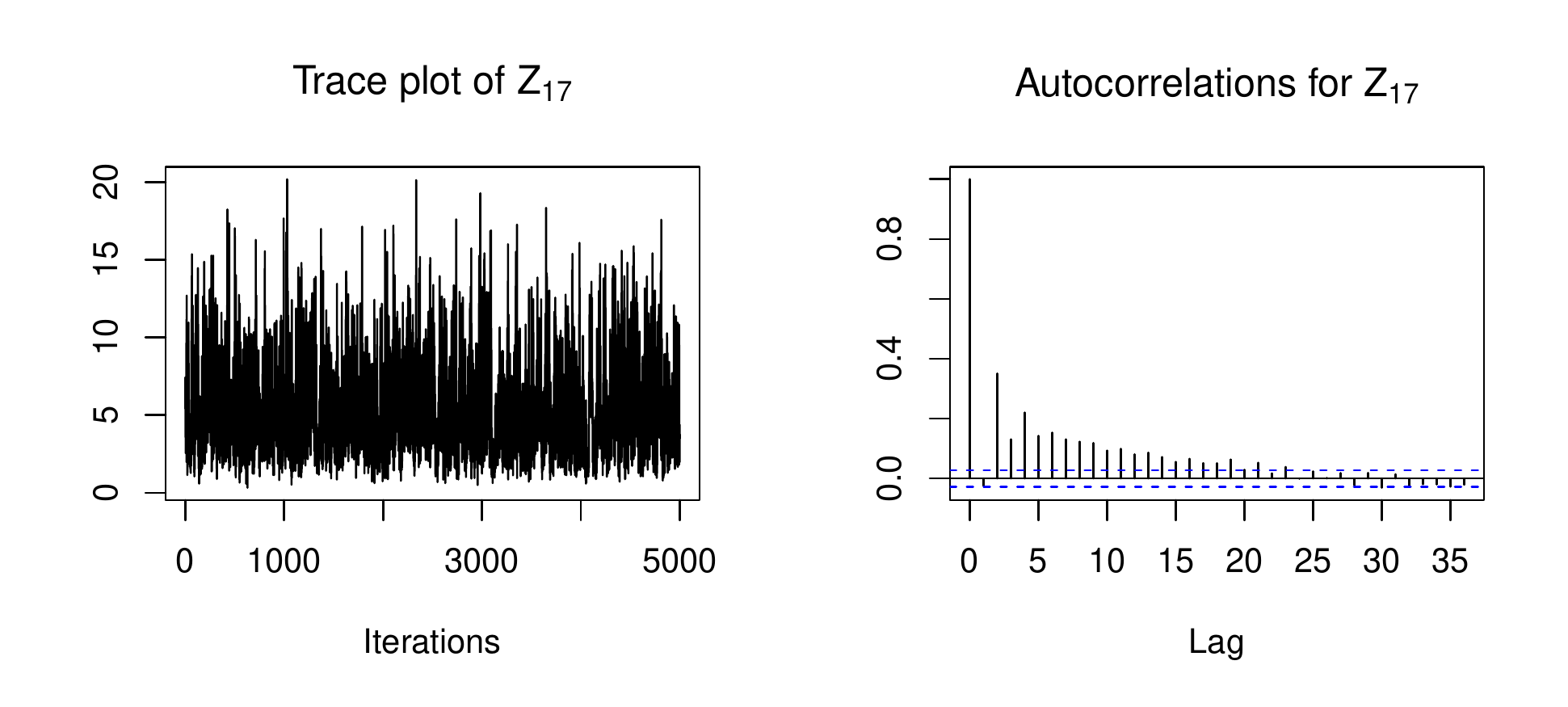}
\caption{Markov chain and autocorrelation plots for the HMC algorithm - $Z_{17}$.}\label{convZ17}
\end{figure}	

\begin{figure}[!htb]
\centering
\includegraphics[scale=0.4]{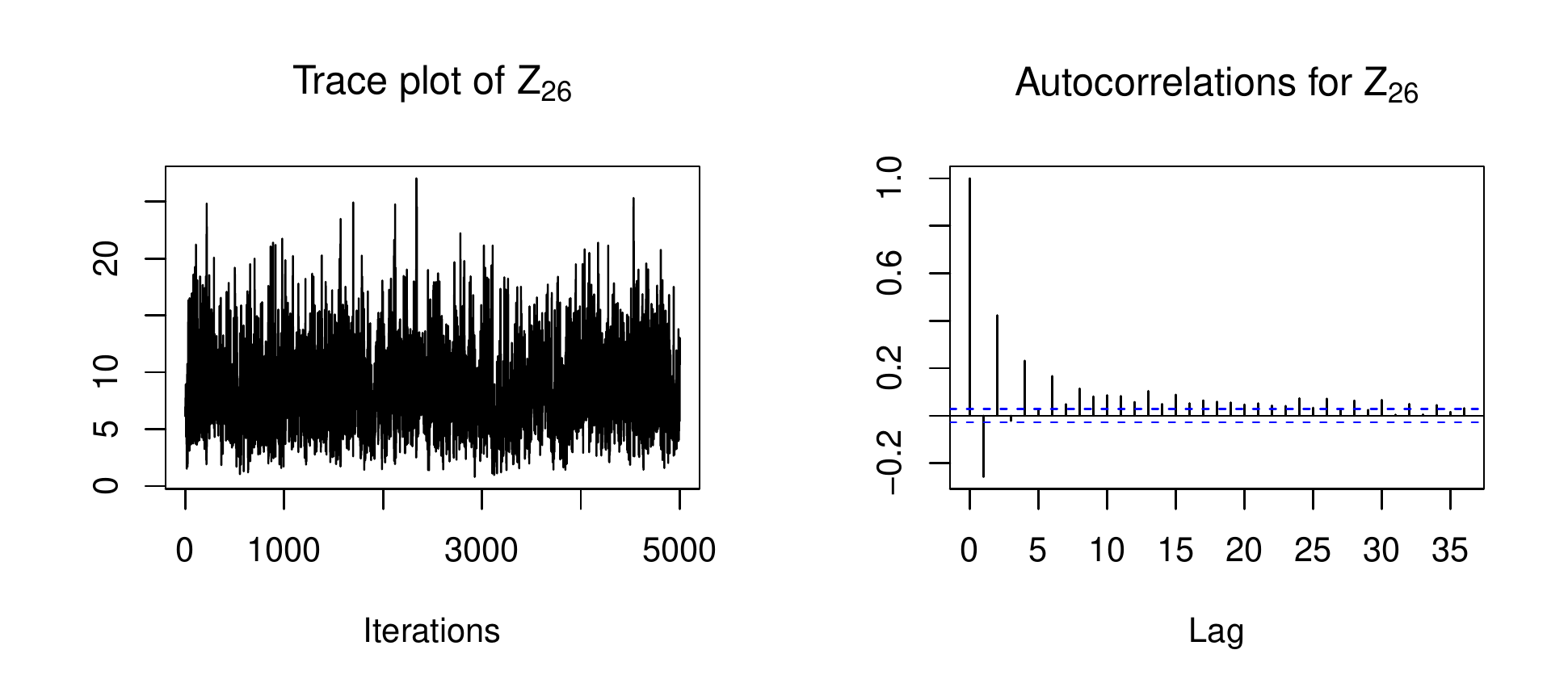}
\caption{Markov chain and autocorrelation plots for the HMC algorithm - $Z_{26}$.}\label{convZ26}
\end{figure}	

\begin{figure}[!htb]
\centering
\includegraphics[scale=0.4]{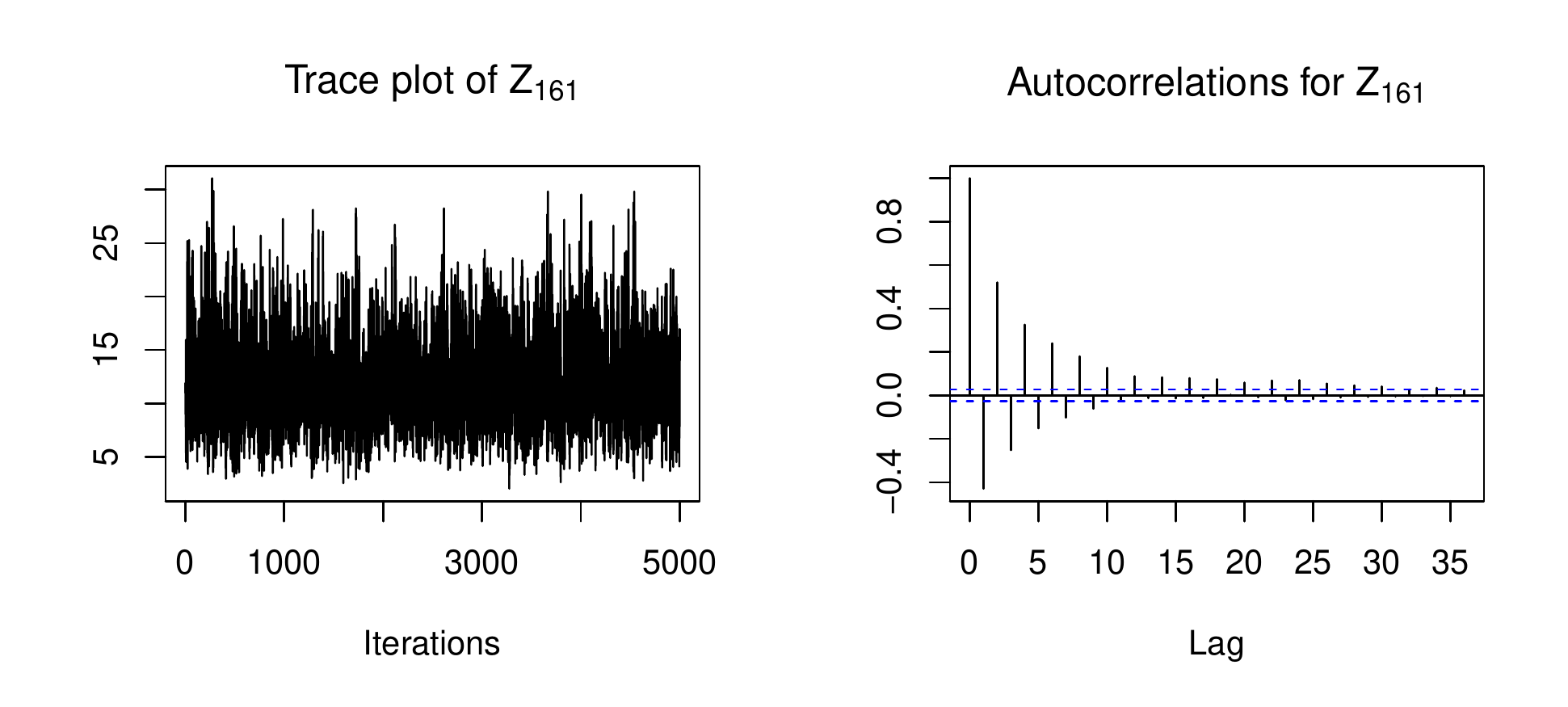}
\caption{Markov chain and autocorrelation plots for the HMC algorithm - $Z_{161}$.}\label{convZ161}
\end{figure}	

\begin{figure}[!htb]
\centering
\includegraphics[scale=0.4]{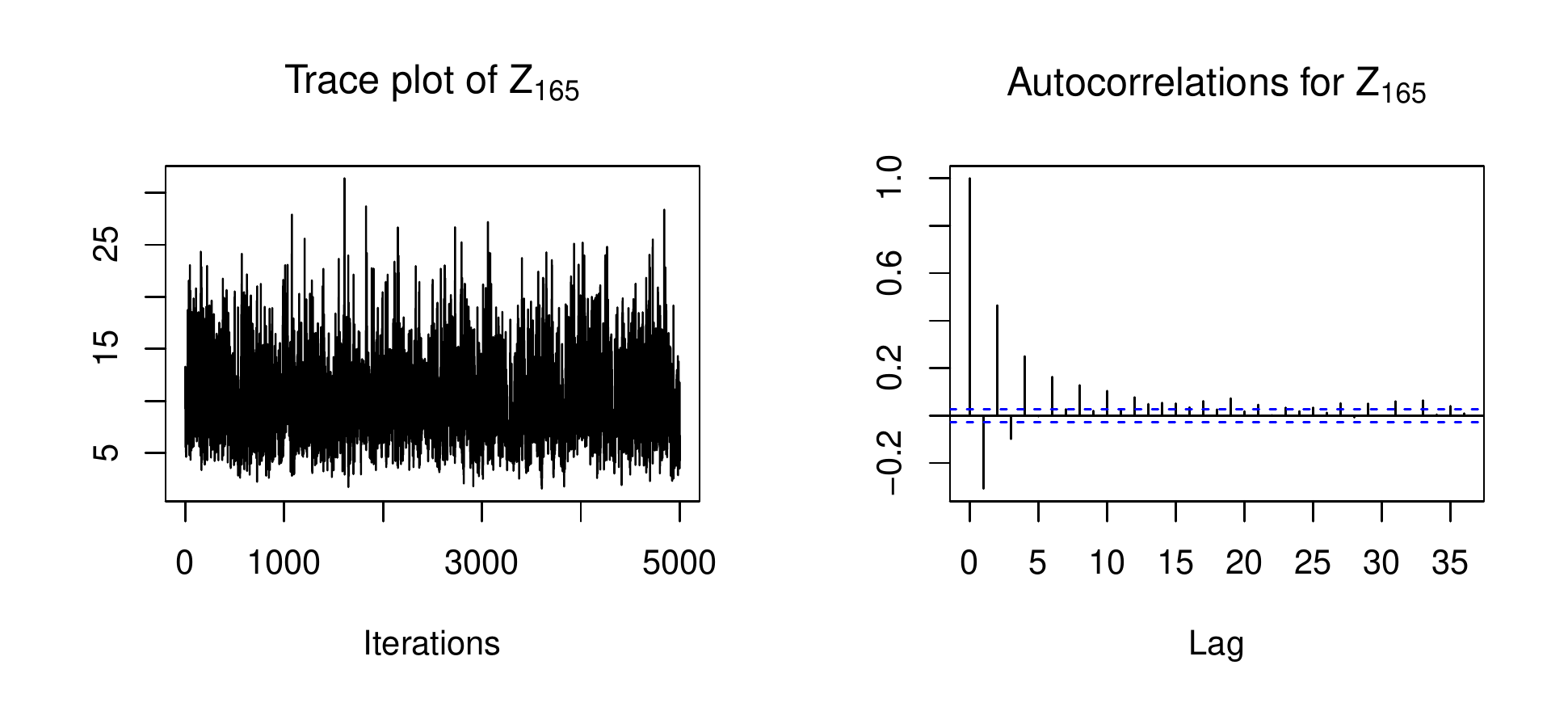}
\caption{Markov chain and autocorrelation plots for the HMC algorithm - $Z_{165}$.}\label{convZ165}
\end{figure}	

\begin{figure}[!htb]
\centering
\includegraphics[scale=0.4]{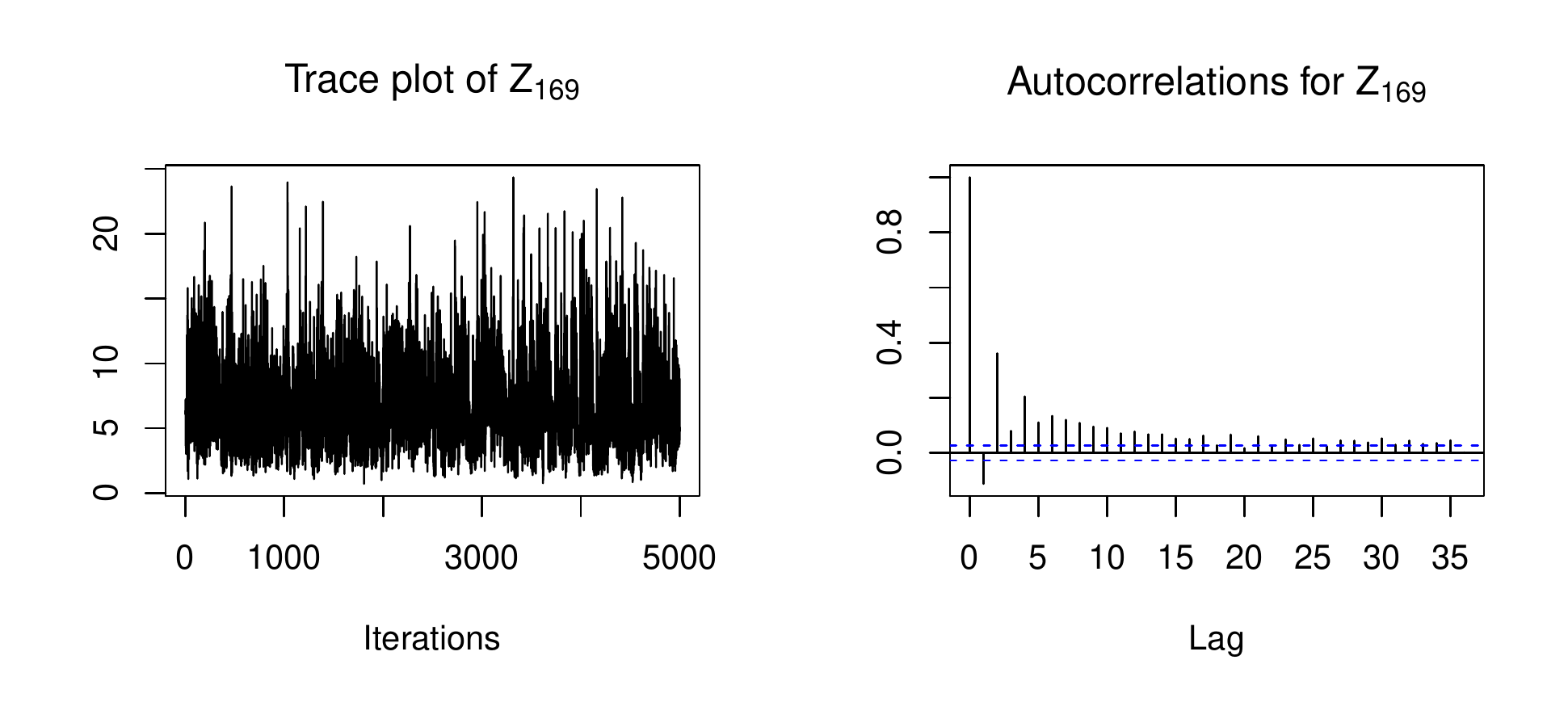}
\caption{Markov chain and autocorrelation plots for the HMC algorithm - $Z_{169}$.}\label{convZ169}
\end{figure}	

\newpage

\section*{Appendix C}\label{chapter:AppendixC}

\begin{table*}[!ht]
\renewcommand{\arraystretch}{1.3}
\caption{Data structure - Observations for {\em m} systems with {\em K} competing risks.}
\centering
\begin{tabular}{cccc}
\hline\hline
System & Competing Risks ($\delta$)& Failure times ($t_{ji}$) & Number of failures ($n_{jq}$)\\ 
\hline\hline
& 1 & $t_{11}, t_{12}, \dots, t_{1n_{11}}$ & $n_{11}$ \\ 
1 & 2 & $t_{11}, t_{12}, \dots, t_{1n_{12}}$ & $n_{12}$ \\ 
& \vdots & \vdots & \vdots  \\ 
& $K$ & $t_{11}, t_{12}, \dots, t_{1n_{1K}}$ & $n_{1K}$  \\ \hline
& \vdots & \vdots & \vdots   \\ 
\vdots & \vdots & \vdots & \vdots   \\ 
& \vdots & \vdots & \vdots  \\ 
&  \vdots & \vdots & \vdots   \\ 
\hline
& 1 & $t_{m1}, t_{m2}, \dots, t_{mn_{m1}}$ & $n_{m1}$ \\ 
m & 2 & $t_{m1}, t_{m2}, \dots, t_{mn_{m2}}$ & $n_{m2}$ \\ 
& \vdots & \vdots & \vdots  \\ 
& $K$ & $t_{m1}, t_{m2}, \dots, t_{mn_{mK}}$ & $n_{mK}$  \\
\hline \hline
\end{tabular}\label{tablemodel2}
\end{table*}

\newpage

\section*{Acknowledgment}
The work is supported by the Brazilian Institutions: EMATER-PAR\'A, CNPq, CAPES and FAPESP.

\ifCLASSOPTIONcaptionsoff
  \newpage
\fi

\bibliographystyle{IEEEtran}

\bibliography{bare_jrnl}

\end{document}